\documentstyle[amssymb,12pt]{article}

\newcommand{\be}{\begin{equation}}
\newcommand{\ee}{\end{equation}}
\newcommand{\ba}{\begin{eqnarray}}
\newcommand{\ea}{\end{eqnarray}}

\begin{document}

\title{{\bf Quantum gravity and spin 1/2 particles effective dynamics}}
\author{{\rm Jorge Alfaro$^a$\thanks{%
jalfaro@fis.puc.cl}, Hugo A. Morales-T\'ecotl$^{bc}$\thanks{
hugo@xanum.uam.mx} \thanks{%
Associate member of Abdus Salam International Centre for Theoretical
Physics, Trieste, Italy.}} \\
and Luis F. Urrutia$^d$\thanks{
urrutia@nuclecu.unam.mx} \\
\\
$^a$Facultad de F\'{\i}sica \\
Pontificia Universidad Cat\'olica de Chile\\
Casilla 306, Santiago 22, Chile\\
$^b$Center for Gravitational Physics and Geometry, Department of Physics \\
The Pennsylvania State University, 104 Davey Lab\\
University Park PA 16802, USA\\
$^c$Departamento de F\'{\i}sica \\
Universidad Aut\'onoma Metropolitana Iztapalapa \\
A.P. 55-534, M\'exico D.F. 09340, M\'exico\\
$^d$Departamento de F\'{\i}sica de Altas Energ\'{\i}as\\
Instituto de Ciencias Nucleares \\
Universidad Nacional Aut\'onoma de M\'exico \\
A.P. 70-543, M\'exico D.F. 04510, M\'exico }
\date{}
\maketitle

\newpage

\begin{abstract}
Quantum gravity phenomenology opens up the possibility of probing 
Planck scale physics. Thus, by exploiting the generic properties that a semiclassical state of the compound system 
fermions plus gravity should have,  an effective dynamics of spin-1/2 particles is obtained within the framework of loop quantum gravity.
 Namely, at length scales much larger than Planck length $\ell_P\sim 10^{-33}$cm and below the wave length of the fermion,  the spin-1/2 dynamics in flat spacetime includes Planck scale corrections. In particular we obtain modified dispersion relations {\em in vacuo} for fermions.
These corrections yield a time of arrival delay of the spin 1/2 particles with respect to a light signal and, in the case of neutrinos, a novel flavor oscillation. 
To detect these effects the corresponding particles must be highly energetic and should travel long distances.  Hence
Neutrino Bursts accompanying Gamma Ray Bursts or ultra high energy cosmic rays could be considered. Remarkably, future neutrino telescopes may be capable to test such effects. This paper provides a detailed account of the calculations and  elaborates on results previously reported in a Letter. These are further amended by introducing a real parameter $\Upsilon$ aimed at encoding our lack of knowledge of scaling properties of the gravitational degrees of freedom.
\end{abstract}

\newpage


\baselineskip=20pt

\section{Introduction}

It is commonly accepted that  quantum gravity should hold at scales near
Planck length $\ell_P:=\sqrt{G_{\rm Newton}\hbar/c^3} \sim 10^{-33}$cm or, equivalently, Planck energy $E_P:={\hbar c}/{\ell_P}\sim 10^{19}$GeV. Accordingly, neither astrophysical observations nor ground experiments were considered in the past as a means to directly reveal any quantum gravity effect but only to test indirect consequences. In recent years however this attitude has changed  on the basis of potentially testable phenomena probing quantum gravity scenarios in which scales combine to lie not far from experimental resolution. Prominent among such phenomena we find {\em in vacuo} dispersion relations for gamma ray astrophysics \cite{HUET,AC,GP,URRU1,FOT,gleiser}, laser-interferometric limits on distance fluctuations \cite{ACLFLUC,NGVDAMLFLUC}, neutrino oscillations \cite{URRU1,BRUSTEIN}, threshold shift in certain high energy physics processes \cite{KIFUNE,ACPIRAN,AMC4,ALF}, CPT violation \cite{CPTELLIS} and clock-comparison experiments in atomic physics\cite{SUDVUUR}. These are the prototypes of  the emerging {\em quantum gravity phenomenology} \cite{ELLIS,AHLU,ACPol,ACEssay,ACINDIA}. 

The present work is aimed at elaborating on effective  corrections to propagation {\em in vacuo}  for spin-1/2 particles in the framework of loop quantum gravity as reported in \cite{URRU1}. This framework also has been used previously in studying light propagation \cite{GP,FOT}.

In essence, the specific structure of spacetime for a given quantum gravity scenario can be probed by matter propagating and interacting there: what  could be considered as flat 
spacetime macroscopically, might produce microscopic imprints of its detailed structure on the interaction of particles. In particular, dispersion relations of propagating matter could exhibit corrections due to such effects. For particles with energy $E<\!\!<E_P$  and momentum $\vec p\,$  the following {\em in vacuo} modified dispersion relations were proposed \cite{AC}:
\begin{eqnarray}
c^2 \, {\vec p}{\,\,}^2 &=& E^2\left(1 + \xi\, \frac{E}{E_{QG}}+ {\cal O}\left(
\frac{E}{E_{QG}}\right)^2 \right)\,,\quad  \label{eq:drel} 
\end{eqnarray}
where $E_{QG}\stackrel{<}{\sim} E_P$ and $\xi\sim 1$. In general, such corrections might behave as $\left(\frac{E}{E_{QG}}\right)^{\Upsilon+1}$, where ${\Upsilon}\geq 0$, namely not necessarily an integer. This possibility has been considered for photons \cite{FOT,THIESAHL} and spin zero particles \cite{THIESAHL}. In this work we show it applies also to spin 1/2 particles. 

According to Eq.(\ref{eq:drel}) the corresponding particle's speed yields a retardation time 
\begin{eqnarray}
\Delta t \approx \xi\, \frac{E}{E_{QG}}\, \frac{L}{c}\,,
\label{eq:Deltat}
\end{eqnarray}
with respect to a speed c signal after traveling a distance $L$. Interestingly, for Gamma Ray Bursts (GRB's) with $E\sim 0.20$MeV,  $L\sim 10^{10}\,$ly and setting $E_{QG}\sim E_P$, the naive estimation Eq. (\ref{eq:Deltat}) gives $\Delta t\sim 0.01$ms, barely two orders of magnitude below the sensitivity $\delta t$ for current observations of GRB's \cite{METZ,BHAT}. Moreover it is  expected to improve this sensitivity in the foreseeable future \cite{meszaros}. 
Using an expression analogous to Eq.(\ref{eq:Deltat}) for the delay of two photons detected with an energy difference $\Delta E$, the observational  bound  $E_{QG}/\xi\geq 4\times 10^{16}$ GeV was established in Ref. \cite{BILLER} by identifying events having $\Delta E=1$ TeV arriving to earth within the time resolution of the measurement $\Delta t =280$ s, from the active galaxy Markarian 421. 

Now, accompanying GRB's there seems to be Neutrino Bursts (NB) in the range $10^{5}-10^{10}$ GeV according to the so-called fireball model  \cite{WAX,VIETRI}. If detected they could provide an excellent means to test quantum gravity effects of the type given by Eq.(\ref{eq:Deltat}) above. Experiments like Neutrino Burst Experiment (NuBE) might detect $\approx$20 events per year of ultra high energy neutrinos ($E>$TeV) coinciding with GRB's \cite{ROY}. Among other experiments aimed at studying ultra high energy cosmic rays including neutrinos we find the OWL-Airwatch project which could detect $\sim 3\times 10^3-10^5$ events ($E>10^{20}$eV) per year  \cite{cline,halzen}. This experiment also can look for time correlations between high energy neutrinos and GRB's. Complementary to the GRB effect, there is the possibility of looking for neutrino oscillation effect as induced by quantum gravity \cite{URRU1}. An analysis
similar to that performed in the case of  atmospheric neutrino oscillations  \cite{BRUSTEIN,SOUTHHAMPTON}, can help to set bounds on the parameters entering the description as we will suggest below.

In summary, astrophysical observations of photons, neutrinos or cosmic rays could make possible to test quantum gravity effects or at  least to restrict quantum gravity theories. 

Indeed, an alternative approach to quantum gravity is based on string theory \cite{REVIEW}. On such a basis  modified dispersion relations of the type (\ref{eq:drel}) have been regained \cite{ELLISETAL}. The main difference in the case of photons is the absence of helicity dependence that is present in the loop quantum gravity approach. As for the case of spin 1/2 particles both approaches seem to agree in its helicity independence to leading order of the corresponding effect \cite{URRU1,ELLISFERM,LAMBIASE}. 

Further comments in order here are: delay time effects for traveling particles have also been considered on a different basis, for example an open system approach \cite{opensystems}. Also, perturbative quantum gravity has been considered to obtain effective dispersion relations \cite{oneloop-effqg}. 

Noticeably, the effect considered thusfar involves a Lorentz symmetry violation \cite{GP,URRU1,ELLISFERM,ELLISETAL}, which seems to be in agreement with some astrophysical and cosmological scenarios \cite{ACPIRAN,ALF,jacobson,MAJOR}.  In this way, these studies naturally overlap with the systematic approach developed by Colladay, Kostelecky and collaborators \cite{colladay} which provides the most general power counting renormalizable extension of the standard model that incorporates both Lorentz and CPT violations. This framework has been widely used to set experimental bounds upon the interactions that produce such violations and the observations performed so far cover a wide
range of experimental settings \cite{BLUHM}. Additional  progress  in establishing  bounds to such symmetry violation can be found in Refs.\cite{gleiser,SUDVUUR,tritium,liberati}.

Finally, it is interesting to emphasize that Planck scale corrections to either particle propagation or interactions need not necessarily imply  violations of Lorentz covariance \cite{PADDY}. Recently the use of non-linear representations of the Lorentz group, leading to what has been called special relativity with two invariant scales, allows for a systematic construction of theories exhibiting these features \cite{SMOLIN}.

The paper is organized as follows. In Section 2 we summarize the basics of loop quantum gravity in the case of fermions plus gravity. Section 3 is devoted to review the regularization of Thiemann for the corresponding Hamiltonian constraint. Section 4  explains general aspects of our approximation scheme whereas Section 5  provides the details of the  calculations. In Section 6  the effective Hamiltonian for non-interacting spin 1/2 particles is obtained. In particular, {\em in vacuo} dispersion relations  are given. Section 7 contains some preliminary estimates of the parameter $\Upsilon$.  To conclude, a summary and discussion of our results is presented in Section 8.

\section{Loop quantum gravity}

\label{LQG} This section provides the basic ingredients of this approach, also known as quantum geometry \cite{volumeop}, which we shall use in the sequel. Among the main results along this
approach one finds: i) well defined geometric operators possessing a discrete spectrum, thus
evidencing discreteness of space \cite{volumeop}, ii) a microscopic account for black
hole entropy \cite{CanonicalBlackHole} and, more recently, hints on quantum avoidance
of a would be classical cosmological singularity \cite{canonicalnosingular}. (For a
review on these topics see for example Ref. \cite{RROV}.)

 To begin with it is assumed that
the spacetime manifold $M$ has topology $\Sigma\times I\!\!R$, with $\Sigma$ a
Riemannian 3-manifold. Here a co-triad $e^i_a$ is defined, with $a,b,c, \dots $ being
spatial tensor indices and $i,j,k,\dots$ being $su(2)$ indices. Thus the
corresponding three-metric is given by $q_{ab}=e^i_ae^i_b$. In addition, a field
$K^i_a$ is defined by
$K_{ab}=sgn [\det(e^j_c)] K^i_a e^i_b$, which is related to the extrinsic curvature $K_{ab}$ of $\Sigma$ .
A canonical pair for the gravitational phase space is $(K^i_a, E_j^b/\kappa)$%
, where $E^a_i=\frac{1}{2}\epsilon^{abc}\epsilon_{ijk}e^j_be^k_c$ and $%
\kappa $ is Newton's constant. It turns out that such a canonical pair yields a
complicated form for the Hamiltonian constraint of general relativity. A convenient
canonical pair, making this constraint polynomial, was introduced by Ashtekar
\cite{Ash}. Nevertheless, two severe difficulties to proceed with the quantization
remained: (i) the implementation of a diffeomorphism covariant regularization for the
density-weight two Hamiltonian constraint hereby obtained and (ii) the extension to
non-compact groups of the diffeomorphism covariant techniques already developed for
gauge theories with compact groups \cite{qdiffgauge}. In fact, the Ashtekar variables
\, ($\, ^{I\!\!\!\!C}\!\!A^i_a=\Gamma^i_a-iK^i_a$, \,\,  $\, iE^a_i/\kappa$) \cite{Ash} , with
$\Gamma^i_a$ being the torsion free connection compatible with $e^i_a$, are complex
valued. Namely the gauge group is $SL(2,I\!\!\!\!C)$, which is non compact.

Some proposals to come to terms with difficulty (ii) were: to consider real
connection variables \cite{Barbero}, to implement a Wick transform %
\cite{WICK} and to define tractable reality constraints \cite{REALITY}. All of these
left open (i).  Thiemann subsequently  proposed to solve (i) and (ii) by incorporating real
connection variables while keeping the density weight one character of the
Hamiltonian constraint. He further provided a quantum version of the theory in the  pure
gravity case, as well as in those cases including the coupling  of  matter to gravity \cite{Thiemann}. His approach is next  reviewed,
since we rely upon it for our analysis of the fermionic case.

As for the fermionic sector a convenient canonical choice is $(\xi,\pi=i\xi^{*} )$. Here $\xi=(\det q)^{1/4}\eta$ is a half density and $\eta$
is a Grassman $SU(2)$ spinor. Half densities are convenient because they do not lead to cumbersome reality conditions at the quantum level and, furthermore, they do not require a complex gravitational connection. However, problems with diffeomorphism covariance can emerge. For this reason a further canonical transformation  in the fermionic sector is necessary in order to dedensitize the Grassman fields through: $\theta(x):=\int_{\Sigma}d^3y\sqrt{\delta(x,y)} \xi(y)$ and 
$\xi(x):= \sum _{y\in\Sigma}\sqrt{\delta(x,y)}\theta(y)$. 

 The pieces of the gravity-spin-1/2 system Hamiltonian constraint read 
\begin{eqnarray}  \label{hmax}
H_{{\rm Einstein}}[N] &=& \int_{\Sigma} d^3x \; N
\frac{1}{\kappa\sqrt{\det q}} {\rm tr} \left( 2\left [%
K_a,K_b]-F_{ab}\right) [E_a,E_b] \right), \nonumber \\
{ H}_{{\rm spin-\frac{1}{2}}}[N] &=& \int_{\Sigma} d^3x \; N E_i{}^a \frac{1}{2\sqrt{det(q)}}
\left(i\pi^T\tau_i{\cal D}_a\xi + {\cal D}_a\left(\pi^T\tau_i\xi\right) + 
\frac{i}{2} K_a{}^i\pi^T\xi + c.c. \right)  \nonumber \\
& & \mbox{}+ \frac{m}{2\hbar}\left(\xi^T( i \sigma^2)\xi +\pi^T ( i
\sigma^2)\pi \right)  \label{eq:h12}
\end{eqnarray}
Here ${\vec \tau}= -\frac{i}{2} {\vec \sigma}$, where $\vec \sigma= \left\{
\sigma^i \right\}$ are the standard Pauli matrices and we have included an explicit mass term. Only one chirality
fermion is used $\eta = \frac{1}{2} ( 1 + \gamma_5) \Psi$, which is
equivalent to have a Dirac spinor $\Psi^T= (\eta^T, \mu^T)$ satisfying the
Majorana condition. The classical configuration space is then ${\cal A}/{\cal G}$ of connections modulo gauge transformations, together with that of the fermionic field.

The quantum arena is given as follows \cite{qdiffgauge}. As in any quantum field
theory, because of the infinite number of degrees of freedom, an enlargement of the
classical configuration space is required. This is far from trivial since the
measures defining the scalar product, which are required to provide a Hilbert space, get
concentrated on distributional fields and hence lie outside the classical configuration
space. The key idea to build up such an enlargement is to make Wilson loop variables
(traces of parallel transport operators) well defined. The
resulting space $\overline{{\cal A}/{\cal G }}$ can be thought of as the limit of
configuration spaces of lattice gauge theories for all possible {\em floating} (i.e. not necessarily rectangular) lattices. Hence, geometric structures on lattice
configuration space are used to implement a geometric structure on $\overline{{\cal
A}/{\cal G}}$. This enables to define a background independent calculus on
it which, in turn, leads to the construction of the relevant measures, the Hilbert space and the regulated operators.

It turns out that the Hilbert space of gauge invariant functions of gravitational and spinor fields is given by 
\cite{Thiemann}
\begin{equation}
{\cal H}_{\rm inv}= L_2\left( \left[\overline{{\cal A}}_{SU(2)} \otimes \overline{\cal S}\right]/\overline {\cal G}, d\mu_{AL}(SU(2))\otimes d\mu_F\right)
\end{equation}
Here $\overline{\cal G}$ denotes the action of $SU(2)$ on all fields at every point of $\Sigma$. 
$\overline S$ denotes the infinite product measurable space $\otimes_x S_x$, $S_x$ being the Grassmann space at the point $x$. $d\mu_F$ is the corresponding measure on $\overline S$. An orthonormal basis using $\mu_F$ can be built up. Let the spinor labels be ordered $A=1,2$. Take $\vec v$ a finite set of mutually different points. For each $v\in \vec v$ denote by $I_v$ the array $(A_1(v)<\cdots<A_k(v))$, with $0\leq k\leq 2$ and $A_j(v)=1,2$ for each $1\leq j\leq k$. Also
set $|I_v|=k$. Thus fermionic vertex functions $F_{\vec v, \vec I}$ can be defined  as $F_{\vec v, \vec I}:= \Pi_{v\in \vec v} F_{v,I}$, $F_{v,I_v}:= \Pi_{j=1}^k \theta_{A_j(v)}(v)$ yielding the desired orthonormal basis.

Gauge invariant objects are spin network states defined as follows. Take $\gamma$ as a piecewise analytic graph with edges $e$ and vertices $v$ which are not necessarily closed. Every edge can be read as outcoming from a vertex by suitably subdividing edges into halves. Given a connection $A_a$ we can compute the holonomy along the edge $e$: 
$h_e(A)$. To each $e$ we also associate a spin $j_e$ corresponding to an irreducible representation of $SU(2)$. In addition, we attach to each vertex an integer label  $n_v$, $0\leq n_v \leq 2$, and a projector $p_v$. For a given $n_v$,
one considers the vector subspace of $Q_v$ (the vector space spanned by holomorphic functions of $\theta_A(v)$, spanned by those vectors $F_{I,v}$ such that $|I|= n(v)$. The projector $p_v$ is a certain $SU(2)$ invariant matrix which projects onto one of the linearly independent trivial representations contained in the decomposition into irreducible representations  of the tensor product consisting of (i) the $n_v$-fold tensor product of fundamental representations of $SU(2)$ associated with the vector subspace of $Q_v$ spanned by the $F_{I,v}, |I|=n(v)$ and (ii) the tensor product of irreducible representations $j_e$ where $e$ runs through the subset of edges of $\gamma$ starting at $v$. The resulting gauge invariant states are denoted by
\begin{equation}
T_{\gamma,[\vec j, \vec n, \vec p]}
\end{equation}
which extend the definition of the matter free case. They form a basis of ${\cal H}_{\rm inv}$. Although not orthonormal it can be transformed into one that it is by suitably decomposing the fermionic dependence into an orthonormal basis of the $Q_v$.

To extract physical information we will further need a state describing a flat
continuous space $\Sigma$ at scales much larger than the Planck length, but not
necessarily so at distances comparable to Planck length itself. States of this kind were introduced under the name of weave \cite{weave} for pure gravity.
 Flat weave
states $|W\rangle$, having a characteristic length ${\cal L}$, were first constructed
by considering collections of Planck scale circles randomly oriented. If one probes
distances $d >\!> {\cal L}$ the continuous flat classical geometry is regained,
while for distances $d <\!< {\cal L}$ the quantum loop structure of space is manifest.
In other words, one expects a behavior of the type $\langle W| {\hat q}_{ab}|W
\rangle= \delta_{ab} + O\left( \frac{\ell_P}{{\cal L}} \right)\,. $ It was soon
realized that such states could not yield a non trivial volume due to the lack of
self intersections \cite{intersecting}. Couples of circles, intersecting at a point,
were also considered as specific models of weaves to overcome this defect
\cite{twocircles}. With the recent advances on the kinematical Hilbert space ${\cal
H}_{{\rm aux}}$ it became clear that all proposed weaves were afflicted by two
undesirable features. First, they are defined to be peaked at a specific (flat or
curved) metric, but not at a connection. This is in
contrast with standard semiclassical states in terms of coherent
states, for example. Second, the known weave states do not belong either to $%
{\cal H}_{{\rm aux}}$ or to a dense subspace of it \cite{gaussweave}. It may be
possible to come to terms with such difficulties by  defining coherent
states for diffeomorphism covariant gauge theories \cite{aei} or by implementing a
genuine statistical geometry \cite{statg}, for instance.  Both approaches  have recently achieved
substantial progress.

 Nonetheless, in order to extract
physics, there is the alternative possibility of using just
the main features that { semiclassical states} should have . Namely, peakedness on both geometry and connection together
with the property that they yield well defined expectation values of physical
operators. An advantage of this alternative is that one may elucidate some physical
consequences before the full fledged semiclassical analysis is  settled down.
Indeed, such alternative may be considered as complementary, in the sense of hinting
at possible features of semiclassical states which could be further elaborated. After
its completion, a rigorous semiclassical treatment should tell us  whether the results
arising from this alternative turn out to hold or not. The weakness of the treatment
resides on its generality, since no detailed features of the { would be
semiclassical states} are used -as opposed, say, to the original weave states- and
hence a set of numerical coefficients cannot be calculated. Evaluating them will be
the task of the rigorous semiclassical treatment.

On top of the {purely gravitational semiclassical states}, a generalization
is required to include matter fields. For our analysis it will just suffice to
 exploit the same aspects of peakedness and well defined expectation
values, extended to include the case of the fermion field. The semiclassical states  here considered will describe flat space and a smooth spin-1/2 field living in
it. Such a state is denoted by $|W,\xi>$ and has a characteristic length ${\cal L}$.  Since no detailed information is used on how the semiclassical
state is constructed in terms of, say, a graph, as opposed to weave states, the present
approach yields results relying only on the following assumptions: (i) peakedness of
the states, (ii) well defined expectation values and iii) existence of a
coarse-grained expansion involving ratios of the relevant scales of the problem: the
Planck length $\ell_P$, the characteristic length ${\cal L}$ and the fermion
wavelength $\lambda$. States fulfilling  such requirements are referred to as {\it would be semiclassical states} in the sequel.

\section{ The regularization}

We will use the regularization method of Thiemann \cite{Thiemann} in the
following. We focus on the first term of $H_{\rm spin-1/2}[N]$, Eq. (\ref{eq:h12}). The other terms can be dealt with similarly or, in the case of $K_a^i$ dependence, along the lines of the purely gravitational sector $H_{\rm Einstein}[N]$ as shown in the sequel. 
The identity $\frac{1}{\kappa} \left\{ A^i_a,V\right\}= 2{sgn}(\det e^j_b) e^i_a$ yields
\begin{equation}  \label{eq:REG1}
H^{(1)}_{\rm spin\frac{1}{2}}[N] = - \frac{i}{2\kappa^2}\int d^3x N(x) \left(\epsilon^{ijk}\epsilon^{abc} 
4\frac{\{A_a^i(x),V(x,\delta)\} \{A_b^j(x),V(x,\delta)\}}{\sqrt{q}(x)} \right)
[\pi^T\tau_k{\cal D}_c\xi -c.c.]  
\end{equation}
where $\delta$ is small and $V(x,\delta)=\int d^3y \; \chi_{\delta}(x,y)\sqrt{q(x)}$ is the volume with respect to the metric $q_{ab}$ of a box centered at $x$. $\chi_{\delta}(x,y) = \prod_{a=1}^3 \theta(\delta/2-|x^a-y^a|)$ is the characteristic function of the box.
Next one changes from $\xi$ to $\theta$ variables and introduces  
\begin{equation}\label{eq:pointsH1}
\sum_x f^a_i (x)(\tau_i {\cal D}_a \theta)_A (x)\bar\theta_A(x) \; ,
\end{equation}
which regularizes (\ref{eq:REG1}) as we show now. Here $f^a_i$ is a real  valued ad$_{SU(2)}$ vector field.
By defining $(\partial_a\theta_A)(x):= \lim_{\epsilon\rightarrow 0}\partial_{x^a}\theta_A^{\epsilon}(x)$, and recalling 
\begin{equation}
\theta_A(x)=\int d^3y\sqrt{\delta(x,y)} \xi_A(y)= \lim_{\epsilon\rightarrow 0}\theta_A^{\epsilon}
\end{equation}
with $\theta_A^{\epsilon} = \int d^3y \frac{\chi_{\epsilon}(x,y)}{\sqrt{\epsilon^3}} \xi_A$, one gets
\begin{equation}
(\partial_{x^a}\theta_A)(x)= \int d^3y \frac{\chi_{\epsilon}(x,y)}{\sqrt{\epsilon^3}} (\partial_{y^a}\theta_A)(y) 
\end{equation}
Now let us divide  $\Sigma$ into boxes of Lebesgue measure $\epsilon^3$ centered at $x_n$ and let 
Eq.(\ref{eq:pointsH1}) be the limit $\epsilon\rightarrow 0$ of 
\begin{equation}
\sum_n f^a_i (x_n)(\tau_i {\cal D}_a \theta^{\epsilon})_A(x_n) \bar\theta^{\epsilon}_A(x_n) \; ,
\end{equation}
which upon expressing $\theta^{\epsilon}$ in terms of $\xi$ becomes
\begin{equation}
\int d^3x \int d^3y \left[ \sum_n f_i^a(x_n)\frac{\chi_{\epsilon}(x,x_n)\chi_{\epsilon}(y,x_n)}{\epsilon^3}\right] 
\left[ \tau_i\partial_a\xi(x) + (\omega_a(x_n)\xi(x))  \right]^A \bar\xi_A \; .
\end{equation}
Notice that in spite of the density weight of $\xi$ no Christoffel connection is needed since it will drop after the h.c part is considered.  In the limit $\epsilon\rightarrow 0$, $\chi_{\epsilon}(x,x_n)/\epsilon^3\rightarrow \delta(x,x_n)$
and $\chi_{\epsilon}(x,x_n)\rightarrow \delta_{y,x_n}$, so that after $x$ integration and sum over $x_n$ the result is
Eq.(\ref{eq:REG1}) where $f_i^a$ has to be properly identified.

To proceed with the quantization one keeps fermionic momenta to the right and replaces $\hat{\bar\theta}^A$ by $\hbar\frac{\partial}{\partial\theta_A}$ in (\ref{eq:pointsH1}). By further multiplying and dividing by $\delta^3$ the factor $\delta^3\sqrt{q(x)}$ in the denominator is changed to $V(x,\delta)$. This can be absorbed into the Poisson brackets already appearing in (\ref{eq:REG1}). When Poisson brackets are replaced by commutators times $1/i\hbar$ one obtains an operator whose action on a cylindrical function $f_\gamma$ associated to a graph $\gamma$ and containing fermionic insertions $\theta_A$ at the vertices $v\in V(\gamma)$ gives
\begin{eqnarray}
\hat{H}^{(1)}_{\rm spin-1/2}[N] f_{\gamma}&=& - \frac{\hbar}{2\ell_P^4} \sum_{v\in V(\gamma)} \sum_x N(x) \epsilon^{ijk} \epsilon^{abc} \; \delta^3 \nonumber\\
&& \mbox{} \times \left[A_a^i(x), \sqrt{\hat V(x,\delta)}\right] \left[A_b^j(x), \sqrt{\hat V(x,\delta)}\right] \nonumber \\
&& \mbox{} \times \left[(\tau_k{\cal D}_c\theta)_A(v)\frac{\partial}{\partial\theta_A(v)} \delta_{x,v}+ h.c.\right] 
f_{\gamma} \; .
\end{eqnarray}
 To complete the regularization an adapted triangulation to $\gamma$ of $\Sigma$ is introduced. Using $h_s(0,\delta) \theta(s(\delta)-\theta(s(0))= \delta\; \dot s ^a(0) ({\cal D}_a)\theta(s(0))$ and re\-plac\-ing the connection operators by ho\-lo\-no\-my op\-er\-a\-tors allows one to absorb the $\delta^3$ factor. Also one replaces $\hat V(v,\delta)\rightarrow \hat V_v$. 
Because of the presence of $\delta_{x,v}$, $\sum_x$ becomes concentrated in the vertices $\sum_{v(\Delta)=v}$.
Hence we obtain
\begin{eqnarray}
\hat{H}^{(1)}_{\rm spin-1/2}[N] &=& - \frac{m_P}{2\ell_P^3} \sum_{v\in V(\gamma)}  N_v  \sum_{v(\Delta)=v}
\epsilon^{ijk} \epsilon^{IJK} \nonumber\\
&& \mbox{} \times {\rm tr} \left( \tau_i h_{s_I(\Delta)}\left[h^{-1}_{s_I(\Delta)}, \sqrt{\hat V_v)}\right]\right)  {\rm tr} \left( \tau_j h_{s_J(\Delta)}\left[h^{-1}_{s_J(\Delta)}, \sqrt{\hat V_v)}\right]\right)\nonumber \\
&& \mbox{} \times \left[\left[(\tau_k h_{s_K(\delta)}\theta)(s_K(\Delta)(\delta))-\theta(v)\right]_A\frac{\partial}{\partial\theta_A(v)} + h.c.\right] \; 
\end{eqnarray}
which is the operator we shall use below.

At this stage we introduce the notation

\begin{equation}  \label{DEFWG}
{\hat w}_{i I \Delta}= {\rm tr}\left( \tau_i h_{s_I(\Delta)} \left[h^{-1}_{s_I(\Delta)},\sqrt{V_v}\right] \right).
\end{equation}


\newpage

\section{General structure of the calculation}

The effective Dirac Hamiltonian ${\rm H}^{F}$ is obtained by considering the
expectation value of the fermionic sector of the quantum Hamiltonian, with respect to $|W,\xi \rangle $. Inside this expectation
value, operators are expanded around all relevant vertices of the
triangulation in powers of the segments $s_{L}^{a}(\Delta )$, having lengths
of the order $\ell _{P}$. In this way, a systematic approximation is giv\-en
involving the scales $\ell _{P}<\!\!<{\cal L}\,<\,\lambda _{D}$. Here $%
\lambda _{D}$ is De Broglie wavelength of the neutrino. Corrections arise at
this level. Let us start by taking the would be semiclassical state (WBSC)\
expectation value of Thiemann's regularized Hamiltonian 
\begin{eqnarray}
{\rm H}^{F}=\langle W,\xi |\hat{H}^{(1)}_{\rm spin-1/2}[N]|W,\xi \rangle =+\frac{%
\hbar }{4\ell _{P}^{4}}\sum_{v\in V(\gamma )}N(v)\frac{8}{E(v)}\epsilon
^{ijk}\epsilon ^{IJK}\times &&  \nonumber \\
\times \mbox{}\left\{ {}\right. \langle W,\xi |\,\left[ \frac{\partial \;}{%
\partial \theta (v)}\tau _{k}h_{s_{K}(\Delta )}{\hat{\theta}}(v+s_{K}(\Delta ))%
\right] \;\hat{w}_{iI\Delta }(v)\hat{w}_{jJ\Delta }(v)|W,\xi \rangle && 
\nonumber \\
\mbox{}-\langle W,\xi |\,\left[ \frac{\partial \;}{\partial \theta (v)}\tau _{k}%
{\hat{\theta}}(v)\right] \,\hat{w}_{iI\Delta }(v)\hat{w}_{jJ\Delta }(v)|W,\xi
\rangle -c.c.\left. {}\right\} &&  \label{EXPW}
\end{eqnarray}%
where $\frac{\partial \;}{\partial \theta (v)}%
\; $ is the fermionic momentum operator .

Our strategy to use the proposed regularization as a computational tool will
be to expand the expression (\ref{EXPW}) around each vertex of the triangulation in powers of  the vectors $s_I(\Delta)$. 
To proceed with the approximation we think of space as made up of boxes of
volume ${\cal L}^{3}$, whose center is denoted by ${\vec{x}}$. Each box
contains a large number of vertices of the graph associated to the WBSC, but is considered
infinitesimal in the scale where the space can be regarded as continuous, so
that we take ${\cal L}^{3}\approx d^{3}x$.

Next  we discuss how to estimate the  average contribution in each box. To begin with we consider
that the volume of the box is 
\begin{equation}
\label{VOLBOX}
d^{3}{\vec{x}}\approx\sum_{v\in {\rm Box}(\vec{x})}\ell _{P}^{3}\;\left( \frac{8}{%
E(v)}\right) 
\end{equation}%
and define the average $<T({\vec{x}})>$ of the quantity $T(v)$ defined in each of the vertices contained in the box,  as 
\begin{equation}
\label{AVE}
d^{3}{\vec{x}}\,\,<T({\vec{x}})>=\sum_{v\in {\rm Box}(\vec{x})}\ell
_{P}^{3}\;\left( \frac{8}{E(v)}\right) \,T(v)
\end{equation}%
The WBSC expectation value of the Hamiltonian (\ref{EXPW}) is of the type 
\begin{eqnarray}
\sum_{v\in V(\gamma )}\frac{8}{E(v)}\langle W,\xi |\hat{F}(v)\hat{G}%
(v)|W,\xi \rangle  &=&\sum_{{\rm Box}(\vec{x})}F(\vec{x})\sum_{v\in {\rm Box}%
(\vec{x})}\ \ell _{P}^{3}\ \frac{8}{E(v)}\langle W,\xi |\frac{1}{\ell
_{P}^{3}}\hat{G}(v)|W,\xi \rangle   \nonumber \\
&=&\int_{\Sigma }F(\vec{x})\;<\frac{1}{\ell^3_P}\,G(\vec{x})>\; d^{3}x
\label{GWEV}
\end{eqnarray}%

Here, ${\hat{F}}(v)$ is a fermionic operator which produces the slowly
varying function $F({\vec{x}})$ within the box; i.e. ${\cal L}<\!\!<\lambda
_{D}$. Furthermore, $<\frac{1}{\ell_P^3}\, G(\vec{x})>$ is the box average of the  rapidly varying WBSC  ex\-pectation values
of  the gravitational operator $\frac{1}{\ell _{P}^{3}}\hat{G}%
(v)$ within the box, whose tensorial and Lie-algebra structure is determined  by the  tensors characterizing the continuum  spacetime we are dealing with, i.e. 
$\stackrel{0}{e}\!{}^{ia},\tau ^{k},\partial _{b},\epsilon ^{cde},\epsilon ^{klm}$, where 
$q^{ab}=\stackrel{0}{e}\!{}^{ia}  \stackrel{0}{e}{}\!_{i}{}^{b}$ is the corresponding classical
3-metric. The order of magnitude of these box averaged quantities  is estimated according to some prescriptions to be specified in the sequel. The method does not provide exact numerical coefficients which can only be obtained from a detailed knowledge of the semiclassical state.

Since we assume the fermionic fields to be slowly-varying functions inside
each box, we demand the following behavior of the fermionic operators inside
the WBSC
\be
\label{FVEV}
\langle W,\xi |....\,{\hat{\theta}}_{B}(v).....\frac{\partial \;}{\partial \theta
^{A}(v)}.....|W,\xi \rangle =\Theta \,\left( ....\xi _{B}(v)....\frac{i}{%
\hbar }\pi _{A}(v)......\right), 
\ee
where the normalization constant $\Theta $ is to be determined in such a way that
the zeroth order approximation reproduces the corresponding classical
kinetic energy term in the Hamiltonian. In this way we have%
\begin{eqnarray}
{\rm H}^{F}=\langle W,\xi |\hat{H}^{(1)}_{\rm spin-1/2}[N]|W,\xi \rangle =+\frac{%
i\Theta }{4\ell _{P}^{4}}\sum_{{\rm Box}(\vec{x})}\sum_{v\in {\rm Box}(\vec{x%
})}N(v)\frac{8}{E(v)}\epsilon ^{ijk}\epsilon ^{IJK}\times  &&  \nonumber \\
\times \mbox{}\left\{ {}\right. \langle W,\xi |\,\left[ \frac{\partial \;}{\partial \theta (v)}\tau
_{k}h_{s_{K}(\Delta )}{\theta }(v+s_{K}(\Delta ))\right] \;\hat{w}_{iI\Delta
}(v)\hat{w}_{jJ\Delta }(v)|W,\xi \rangle  &&  \nonumber \\
\mbox{}-\langle W,\xi |\,\left[ \frac{\partial \;}{\partial \theta (v)}\tau _{k}{\theta }(v)\right] \,\hat{w}%
_{iI\Delta }(v)\hat{w}_{jJ\Delta }(v)|W,\xi \rangle -c.c.\left. {}\right\} 
&&
\label{HWEV1}
\end{eqnarray}%
In the above equation we have written only the fermions in its classical version and we have
kept them inside the WBSC expectation value just in order to be able to write the above
expression in a more compact way. The holonomy $h_{s_{K}(\Delta )}$ still
contains the gravitational connection. In
order to have a convenient bookkeeping of the terms involved, we write Eq.(\ref{HWEV1}) as
\begin{eqnarray}
{\rm H}^{F} &=&+\frac{i\Theta }{4\ell _{P}^{4}}\sum_{{\rm Box}(\vec{x}%
)}\sum_{v\in {\rm Box}(\vec{x})}\;N(v)\frac{8}{E(v)}\epsilon ^{ijk}\epsilon
^{IJK}\times \;\;\;  \nonumber \\
&&\times \,\,\{\langle W,\xi |\,\left[ \frac{\partial \;}{\partial \theta (v)}\tau _{k}\left( \left(
s_{K}^{a}\hat{\nabla}_{a}(v)\right) +\frac{1}{2}\left( s_{K}^{a}\hat{\nabla}%
_{a}(v)\right) \left( s_{K}^{b}\hat{\nabla}_{b}(v)\right) +....\right) {\theta }%
(v)\;\right]   \nonumber \\
&&\times \,\,\hat{w}_{iI\Delta }(v)\hat{w}_{jJ\Delta }(v)|W,\xi \rangle
-c.c.\}
\label{HWEV2}
\end{eqnarray}
where the derivative $\hat{\nabla}_a $ includes the
covariant derivative $\stackrel{0}{\nabla}_{a}(\vec{x})$, which is compatible
with the classical metric that we are considering and a piece $\hat{A}_{a}^{i}(v)\;$ producing the quantum corrections associated with the
beginning of the continuum at the scale ${\cal L}$ . Upon taking WBSC expectation value we can make the following substitution
\begin{eqnarray}
\hat{\nabla}_{a}(v){\theta }(v) &\longrightarrow&\left( \stackrel{0}{\nabla}_{a}(\vec{x})+\hat{A}%
_{a}^{i}(v)\tau _{i}\right) {\xi }(\vec{x}) 
\end{eqnarray}%
From now on we restrict to a  continuum flat  metric, in such a way that%
\begin{equation}
\stackrel{0}{\nabla}_{a}(\vec{x})=\partial _{a} \;.
\end{equation}%
The partial derivative does not change inside each box, which is consistent with the idea that each box is supposed to represent a point of the continuum. In this way
\begin{eqnarray}
\label{HWEV3}
{\rm H}^{F} &=&+\frac{i\Theta }{4\ell _{P}^{4}}\sum_{{\rm Box}(\vec{x}%
)}\sum_{v\in {\rm Box}(\vec{x})}\;N(v)\frac{8}{E(v)}\epsilon ^{ijk}\epsilon
^{IJK}\times \;\;\;  \nonumber \\
&&\times \,\,\{\langle W,\xi |\,\left[ \frac{\partial \;}{\partial \theta (v)}\tau _{k}\left( s_{K}^{a}\hat{%
\nabla}_{a}(v)+\frac{1}{2}s_{K}^{a}s_{K}^{b}\hat{\nabla}_{a}(v)\hat{\nabla}%
_{b}(v)+....\right) {\theta }(v)\;\right]   \nonumber \\
&&\times \,\,\hat{w}_{iI\Delta }(v)\hat{w}_{jJ\Delta }(v)|W,\xi \rangle
-c.c.\}
\end{eqnarray}%
Our problem now is to parameterize and to estimate the general  spinor \begin{eqnarray}
\label{SBOX}
Y_{(n)}^{k}(v) &=&\epsilon ^{ijk}\epsilon ^{IJK}\langle W,\xi
|s_{K}^{a_{1}}s_{K}^{a_{2}}....s_{K}^{a_{n}}\hat{\nabla}_{a_{1}}(v)\hat{%
\nabla}_{a_{2}}(v)...\hat{\nabla}_{a_{n}}(v){\theta }(v)\, \nonumber \\
&&\times \hat{w}_{iI\Delta }(v)\hat{w}_{jJ\Delta }(v)|W,\xi \rangle 
\end{eqnarray}%
in  a given box.
Here $n$ denotes the number of  covariant derivatives in the expression. Using the above we can write
\be
\label{HWEV4}
{\rm H}^{F}=+\frac{i\Theta }{4\ell _{P}^{4}}\sum_{n=1,2,...}\sum_{{\rm Box}(%
\vec{x})}\sum_{v\in {\rm Box}(\vec{x})}\;\left( N(v)\frac{8}{E(v)}\right)
\pi (v)\tau _{k}Y_{(n)}^{k}(v)-c.c.\;\;
\ee
Before calculating the corresponding box averaged contributions let us obtain the expressions for some of the first quantities $Y_{(n)}^{k}(v)$.  A direct calculation shows 
\begin{equation}
\label{Y1}
Y_{(1)}^{k}(v)=s_{K}^{a}(v)\left( X^{kK}(v)\partial _{a}{\xi }(\vec{x}%
)+\;X_{a}^{m\;kK}(v)\tau _{m}\right) {\xi}(\vec{x}),
\end{equation}%
\begin{eqnarray}
\label{Y2}
Y_{(2)}^{k}(v) &=&\;s_{K}^{a}(v)s_{K}^{b}(v)X^{kK}(v)\;\partial _{a}\partial
_{b}{\xi }(\vec{x})+\;  \nonumber \\
&&+s_{K}^{a}(v)s_{K}^{b}(v)X_{a}^{m\;kK}(v)\tau _{m}2\partial _{b}{\xi }(%
\vec{x})  \nonumber \\
&&+s_{K}^{a}(v)s_{K}^{b}(v)X_{ab}^{mn\;kK}(v)\tau _{m}\tau _{n}{\xi}(\vec{x}%
),
\end{eqnarray}%
\begin{eqnarray*}
\label{Y3}
Y_{(3)}^{k}(v) &=&s_{K}^{a}s_{K}^{b}s_{K}^{c}\;\left( X^{kK}\;(v)\partial
_{a}\partial _{b}\partial _{c}+3X_{a}^{m\;kK}(v)\tau _{m}\partial
_{b}\partial _{c}+\right) {\xi }(\vec{x}) \\
&&+s_{K}^{a}s_{K}^{b}s_{K}^{c}\;\left( 3X_{ab}^{mn\;kK}(v)\tau _{m}\tau
_{n}\partial _{c}+X_{abc}^{mnp\;kK}(v)\tau _{m}\tau _{n}\tau _{p}\right) {%
\xi }(\vec{x}),
\end{eqnarray*}
where we have explicitly enforced the symmetry implied by the factors$%
\;s_{K}^{a}s_{K}^{b}s_{K}^{c}....\;$ in Eq.(\ref{SBOX}). The  gravitational quantities, which  are
rapidly varying inside the box are%
\begin{eqnarray*}
X^{kK}(v) &=&\epsilon ^{ijk}\epsilon ^{IJK}\langle W,\xi |\,\hat{w}%
_{iI\Delta }(v)\hat{w}_{jJ\Delta }(v)|W,\xi \rangle  \\
X_{a}^{m\;kK}(v) &=&\epsilon ^{ijk}\epsilon ^{IJK}\langle W,\xi |\hat{A}%
_{a}^{m}(v)\hat{w}_{iI\Delta }(v)\hat{w}_{jJ\Delta }(v)|W,\xi \rangle  \\
X_{ab}^{mn\;kK}(v) &=&\epsilon ^{ijk}\epsilon ^{IJK}\langle W,\xi |\hat{A}%
_{a}^{m}(v)\hat{A}_{b}^{n}(v)\hat{w}_{iI\Delta }(v)\hat{w}_{jJ\Delta
}(v)|W,\xi \rangle  \\
X_{abc}^{mnp\;kK}(v) &=&\epsilon ^{ijk}\epsilon ^{IJK}\langle W,\xi |\hat{A}%
_{a}^{m}(v)\hat{A}_{b}^{m}(v)\hat{A}_{c}^{p}(v)\hat{w}_{iI\Delta }(v)\hat{w}%
_{jJ\Delta }(v)|W,\xi \rangle
\label{GRAVQB} 
\end{eqnarray*}%
Next we write the corresponding  general expressions
\begin{eqnarray}
Y_{(n)}^{k}(v) &=&\left( s_{K}^{a_{1}}s_{K}^{a_{2}}...s_{K}^{a_{l}}\right)
\left( s_{K}^{a_{l+1}}s_{K}^{a_{l+2}}...s_{K}^{a_{n}}\right)
\;\sum_{l=n,n-1,....1,0}\left(\frac{}{} _{\, l}^{n}\right)
\;X_{a_{l+1}a_{l+2}....a_{n}}^{m_{l+1}m_{l+2}..m_{n}\;kK}\;(v) \nonumber \\
&&\times \tau _{m_{l+1}}\tau _{m_{l+2}}...\tau _{m_{n}}\partial
_{a_{1}}\partial _{a_{1}}....\partial _{a_{l}}{\xi }(\vec{x})
\label{YGEN}
\end{eqnarray}%
with
\begin{eqnarray}
\label{XGEN}
X_{a_{l+1}a_{l+2}....a_{n}}^{m_{l+1}m_{l+2}..m_{n}\;kK}(v) &=&\epsilon
^{ijk}\epsilon ^{IJK}\langle W,\xi |\hat{A}_{a_{l+1}}^{m_{l+1}}(v)\hat{A}%
_{a_{l+2}}^{m_{l+2}}(v)...\hat{A}_{a_{n}}^{m_{n}}(v)\hat{w}_{iI\Delta }(v)\hat{w}_{jJ\Delta
}(v)|W,\xi \rangle. \nonumber \\ 
\end{eqnarray}%
In the above expressions the partition $(1,2,...l)\;$ and $(l+1,$ $l+2,\;....n\;)$ is made in such a way that the indices of the first set  count the number of partial derivatives acting upon the
fermion. The indices of the second set count the number of $\tau$-matrices involved
which is equal to the number of additional connections, i.e. besides those contained in the combination $\hat{w}_{iI\Delta }(v)\hat{w}_{jJ\Delta
}(v)$, appearing in Eq.(\ref{XGEN}).

The box-averaged expression for the Hamiltonian reduces then to
\be
{\rm H}^{F}=\frac{i\Theta }{4\ell _{P}^{7}}\sum_{n=1,2,...}\int d^{3}{\vec{x%
}}\;\pi ({\vec{x}})\;\tau _{k}<Y_{(n)}^{k}({\vec{x}}%
)>-c.c.\;\;
\ee
In more detail this is 
\begin{eqnarray}
{\rm H}^{F} &=&+\frac{i\Theta }{4\ell _{P}^{7}}\sum_{n=1,2,...}\int d^{3}{%
\vec{x}}\;\pi ({\vec{x}})\;\tau _{k}\times \nonumber \\
&<&\left(
s_{K}^{a_{1}}s_{K}^{a_{2}}...s_{K}^{a_{l}}\right) \left(
s_{K}^{a_{l+1}}s_{K}^{a_{l+2}}...s_{K}^{a_{n}}\right)
\sum_{l=n,n-1,....1,0}\left(\frac{}{} _{\, l}^{n}\right)
\;X_{a_{l+1}a_{l+2}....a_{n}}^{m_{l+1}m_{l+2}..m_{n}\;kK}\;(v)> \nonumber \\
&&\;\tau _{m_{l+1}}\tau _{m_{l+2}}...\tau _{m_{n}}\partial _{a_{1}}\partial
_{a_{1}}....\partial _{a_{l}}{\xi }(\vec{x}).
\label{HWEVF1}
\end{eqnarray}%
Here it is convenient to define the auxiliary quantities%
\begin{eqnarray}
\label{AUXT}
T^{a_{1}a_{2}....a_{l}m_{l+1}m_{l+2}.......m_{n}\;k}\; &=&\frac{\Theta }{%
4\ell _{P}^{7}}<\left(
s_{K}^{a_{1}}s_{K}^{a_{2}}...s_{K}^{a_{l}}\right) \left(
s_{K}^{a_{l+1}}s_{K}^{a_{l+2}}...s_{K}^{a_{n}}\right) \, X_{a_{l+1}a_{l+2}....a_{n}}^{m_{l+1}m_{l+2}..m_{n}\;kK}\;(v) >\nonumber \\
\end{eqnarray}%
in such a way that 
\ba
\label{HWEVF2}
{\rm H}^{F}&=&\sum_{n=1,2,...}\sum_{l=n,n-1,....1,0}\left(\frac{}{} _{\,l}^{n}\right)
\int d^{3}{\vec{x}}\;i\pi ({\vec{x}})\;\tau
_{k}\;\;T^{a_{1}a_{2}....a_{l}m_{l+1}m_{l+2}.......m_{n}}\,\tau _{m_{l+1}}\tau _{m_{l+2}}...\tau _{m_{n}}\nonumber\\
&&\qquad \qquad \qquad \qquad\qquad\qquad 
\qquad \qquad\qquad\qquad \times\partial
_{a_{1}}\partial _{a_{1}}....\partial _{a_{l}}{\xi }(\vec{x})\nonumber \\
\ea
\be
{\rm H}^{F}=:\sum_{n=1,2,...}\sum_{l=n,n-1,....1,0}{\rm H}_{nl}^{F}
\ee
The box-averages $<F(v)>$ are  subsequently estimated using the
corresponding dimensions in terms of the available tensors in flat space 
\begin{equation}
\stackrel{0}{e}\!{}^{ia}=\delta ^{ia},\quad \tau ^{k},\quad \partial _{b},\quad
\epsilon ^{cde},\quad \epsilon ^{klm},\quad q^{ab}=\delta ^{ab},\quad \delta
^{ij},\quad \delta _{b}^{a}
\end{equation}%
in a manner described in the next section. In this way we are demanding
covariance under rotations at the scale ${\cal L}$. As the final input in  our prescription we impose that when
averaging inside each box, the order of magnitude of the corresponding
expectation values of the gravitational operators are estimated according to 
\begin{equation}
\label{RWEV}
\langle W,\,\xi \,|\dots ,A_{ia},\,\dots ,\sqrt{V_{v}},\dots |W,\,\xi
\,\rangle \approx \dots \frac{1}{{\cal L}}\left( \frac{\ell _{P}}{{\cal L}}%
\right) ^{\Upsilon }\stackrel{0}{e}_{ia},\,\dots ,\,\ell _{P}^{3/2}\dots \,.,
\label{Ascale}
\end{equation}%
respectively. 

In previous work \cite{URRU1} we have set $\Upsilon =0$ on
the basis that the coarse graining approximation does not allow for the
continuum connection to be probed below $\frac{1}{{\cal L}}$. On the other hand, by
adopting say kinematical coherent states for representing semiclassical
states, one would set $\Upsilon =1$ for two reasons: first, in the limit $%
\hbar \rightarrow 0 \, \, (\ell_P\rightarrow 0)$ (\ref{Ascale}) yields just zero, in agreement with a
flat connection, and, second, such an scaling would saturate the Heisenberg
uncertainty relation. Nonetheless, physical semiclassical states may lead to
a leading order contribution with $\Upsilon \neq 0,1$, thus we leave it open
here. Now we have all the ingredients to perform the expansion in powers of $%
s$.

Let us consider now the  contribution arising from the  gravitational
operators contained  in $\hat{w}_{iI\Delta }(v)$ of Eq.(\ref{XGEN}). We choose the
parameterization  
\begin{equation}
\hat{w}_{iI\Delta }(v)=s_{I}{}^{a}\ w_{ia}+s_{I}{}^{a}\ s_{I}{}^{b}\
w_{iab}+\dots 
\end{equation}%
leading to 
\begin{equation}
\hat{w}_{iI\Delta }(v)\ \hat{w}_{jJ\Delta }(v)=s_{I}{}^{a}\ s_{J}{}^{b}\
w_{ia}\ w_{jb}+s_{I}{}^{a}\ s_{J}{}^{b}\ s_{J}{}^{c}\ w_{ia}\
w_{jbc}+s_{I}{}^{a}\ s_{I}{}^{b}\ s_{J}{}^{c}\ w_{iab}\ w_{jc}+O(s^{4}),
\end{equation}%
where $w_{ia}$ and $w_{iab}$, which are independent of $s$, need to be
calculated explicitly. The product of ${\hat{w}}$ starts quadratically in $s$%
.

To this end we will need the $\tau$-algebra, recalling that $\tau_k= -\frac{i%
}{2}\sigma_k$, with $\sigma_k$ being the standard Pauli matrices. We have 
\begin{eqnarray}  \label{TAUA}
\tau_i \tau_j&=& -\frac{1}{4} \delta_{ij} + \frac{1}{2} \epsilon_{ijk}
\tau_k,  \nonumber \\
\tau_i \tau_j \tau_k&=&-\frac{1}{8} \epsilon_{ijk} - \frac{1}{4} \delta_{ij}
\tau_k + \frac{1}{4} \delta_{ik} \tau_j - \frac{1}{4} \delta_{jk} \tau_i,\nonumber\\
\epsilon ^{cab}\tau _{a}\tau _{b} &=& \frac{1}{2}\epsilon ^{cab}\epsilon
_{abm}\tau _{m}=\tau _{c}\nonumber\\
\tau _{k}\tau _{a}\tau _{k}&=&
\frac{1}{4}\tau _{a}
\end{eqnarray}
From the definition (\ref{DEFWG}) we obtain 
\begin{equation}
{\hat w}_{i I \Delta}= Tr \left(\tau_i(\sqrt{V}- h_{s_I} \sqrt{V}%
h^{-1}_{s_I}) \right)
\end{equation}
Up to second order in $s$, we have 
\begin{eqnarray}
{\hat w}_{i I \Delta}&=& -Tr \left(\tau_i([A_I ,\sqrt{V}] + \frac{1}{2} [A_I,
[A_I, \sqrt{V}]] + \dots )\right), \quad A_I=s_I{}^a \,A_{ia} \,\tau^i  \nonumber \\
&=& s_I{}^a \ \frac{1}{2} [A_{ia}, \sqrt{V}] + s_I{}^a \ s_I{}^b \ \frac{1}{8%
} \epsilon_{ikl}\ [ A_{k a}, [ A_{l b}, \sqrt{V}]] + \dots,
\end{eqnarray}
which finally allows to identify 
\begin{eqnarray}
w_{ia}= \frac{1}{2} [A_{ia}, \sqrt{V}], \qquad w_{i a b}=\frac{1}{8}
\epsilon_{ikl} \ [ A_{k a}, [ A_{l b}, \sqrt{V}]].
\end{eqnarray}

The scaling properties under the semiclassical expectation value of the
above gravitational operators is 
\begin{equation}
\langle W\,\xi |\dots w_{i\,a_{1}\dots a_{n}}\dots |W\,\xi \rangle
\rightarrow \frac{\ell _{P}{}^{3/2}}{{\cal L}^{n}}\left( \frac{\ell _{P}}{%
{\cal L}}\right) ^{n\Upsilon }.  \label{Ascale1}
\end{equation}

\section{The calculation}

The detailed calculation of the correction terms is performed according to the following
prescription: first we set $\Upsilon=0$, then we consider an expansion to order  $\ell _{P}^{2}$ and finally we reintroduce the non-zero value for $\Upsilon$ in  the corresponding
terms. 

Let us recall the general expressions from last section%
\begin{eqnarray}
{\rm H}_{nl}^{F} &=&i\left( _{l}^{n}\right) \int d^{3}{\vec{x}}\;\pi ({\vec{x%
}})\;\tau _{k}\;\;T^{a_{1}a_{2}....a_{l}m_{l+1}m_{l+2}.......m_{n}\;k}\; \nonumber \\
&&\mbox{} \times \tau _{m_{l+1}}\tau _{m_{l+2}}...\tau _{m_{n}}\partial
_{a_{1}}\partial _{a_{1}}....\partial _{a_{l}}{\xi }(\vec{x})
\end{eqnarray}%
\begin{eqnarray}
T^{a_{1}a_{2}....a_{l}m_{l+1}m_{l+2}.......m_{n}\;k}\; &=&\frac{\Theta }{%
4\ell _{P}^{7}}< \left( s_{K}^{a_{1}}s_{K}^{a_{2}}...s_{K}^{a_{l}}\right)
\left( s_{K}^{a_{l+1}}s_{K}^{a_{l+2}}...s_{K}^{a_{n}}\right) \nonumber \\
&&\mbox{} \times \; X_{a_{l+1}a_{l+2}....a_{n}}^{m_{l+1}m_{l+2}..m_{n}\;kK}\;(v) > \end{eqnarray}%
\begin{eqnarray}
X_{a_{l+1}a_{l+2}....a_{n}}^{m_{l+1}m_{l+2}..m_{n}\;kK}(v) &=&\epsilon
^{ijk}\epsilon ^{IJK}\langle W,\xi |\hat{A}_{a_{l+1}}^{m_{l+1}}(v)\hat{A}%
_{a_{l+2}}^{m_{l+2}}(v)... \nonumber \\
&&\mbox{} \times \hat{A}_{a_{n}}^{m_{n}}(v)\hat{w}_{iI\Delta }(v)\hat{w}_{jJ\Delta
}(v)|W,\xi \rangle
\end{eqnarray}%
The partition is made of the number $n\;$such that we have $l\;\;$%
derivatives and $(n-l)$ $\tau$-matrices, all arising from the presence of $n$
covariant derivatives in Eq.(\ref{HWEV3}). We recall that 
\begin{eqnarray}
\label{EXPWW}
\hat{w}_{iI\Delta }(v)\ \hat{w}_{jJ\Delta }(v) &=&\left( s_{I}{}^{a}\
s_{J}{}^{b}\ w_{ia}\ w_{jb}\right) +\left( s_{I}{}^{a}\ s_{J}{}^{b}\
s_{J}{}^{c}\ w_{ia}\ w_{jbc}+s_{I}{}^{a}\ s_{I}{}^{b}\ s_{J}{}^{c}\ w_{iab}\
w_{jc}\right)  \nonumber \\
&&+O(s^{4}),  \nonumber \\
&=&\left( \hat{w}_{iI\Delta }(v)\ \hat{w}_{jJ\Delta }(v)\right) _{2}+\left( 
\hat{w}_{iI\Delta }(v)\ \hat{w}_{jJ\Delta }(v)\right) _{3}+\nonumber \\
&&...+\left( 
\hat{w}_{iI\Delta }(v)\ \hat{w}_{jJ\Delta }(v)\right) _{N}+...,
\end{eqnarray}%
where $N$\ counts the powers of $s$ in the term $\hat{w}_{iI\Delta }(v)\ 
\hat{w}_{jJ\Delta }(v)$. 
Under the semiclassical state we can
estimate the contribution of each term in  Eq.(\ref{EXPWW}) as 
\be
\langle W,\xi |.....\left( \hat{w}_{iI\Delta }(v)\hat{w}_{jJ\Delta
}(v)\right) _{N}....|W,\xi \rangle =\ell _{P}^{3}\left( \frac{\ell _{P}}{%
{\cal L}}\right) ^{N(1+\Upsilon )}, 
\ee
where we have used Eqs.(\ref{Ascale}) and (\ref{Ascale1}).

\subsection{ The case $n=1$}

Here we have two possibilities which we consider separately: $(n=1,$ $l=1)$
and $(n=1,$ $l=0).$

\subsubsection{Case $n=1,$ $l=1$}

\begin{equation}
{\rm H}_{11}^{F}=i\int d^{3}{\vec{x}}\;\pi ({\vec{x}})\;\tau
_{k}\;\;T_{11}^{a\;k}\;\partial _{a}{\xi }(\vec{x})
\end{equation}
\begin{equation}
T_{11}^{a\;k}\;=\frac{\Theta }{4\ell _{P}^{7}}< s_{K}^{a}\;X^{\;kK}\;(v)>
\end{equation}
\begin{equation}
X^{\;kK}(v)=\epsilon ^{ijk}\epsilon ^{IJK}\langle W,\xi |\hat{w}_{iI\Delta
}(v)\hat{w}_{jJ\Delta }(v)|W,\xi \rangle
\end{equation}
We present this first calculation in some detail since it sets the stage for
all the remaining estimates 
\begin{eqnarray}
T_{11}^{a\;k}\; &=&\frac{\Theta }{4\ell _{P}^{7}}\,< s_{K}^{a}\;\epsilon
^{ijk}\epsilon ^{IJK}\langle W,\xi |\hat{w}_{iI\Delta }(v)\hat{w}_{jJ\Delta
}(v)|W,\xi \rangle \,> \nonumber \\
&=&\delta ^{ak}\frac{\Theta }{4\ell _{P}^{7}}\ell _{P}\sum_{N=2,3,}\vartheta
_{11N}\ell _{P}^{3}\left( \frac{\ell _{P}}{{\cal L}}\right) ^{N(1+\Upsilon )}
\nonumber \\
&=&\delta ^{ak}\frac{\Theta }{4\ell _{P}^{3}}\left( \vartheta _{112}\left( 
\frac{\ell _{P}}{{\cal L}}\right) ^{2(1+\Upsilon )}+\vartheta _{113}\left( 
\frac{\ell _{P}}{{\cal L}}\right) ^{3(1+\Upsilon )}+....\right)  \nonumber \\
T_{11}^{a\;k} &=&\delta ^{ak}\frac{\Theta }{4\ell _{P}^{3}}\left( \frac{\ell
_{P}}{{\cal L}}\right) ^{2(1+\Upsilon )}\left( \vartheta _{112}+\vartheta
_{113}\left( \frac{\ell _{P}}{{\cal L}}\right) ^{(1+\Upsilon )}....\right)
\end{eqnarray}%
The notation is $\vartheta_{nlN}$ and these numbers are assumed to be of order one.
Choosing  $\vartheta _{112}=1$ we are
left with%
\begin{eqnarray}
T_{11}^{a\;k} &=&\delta ^{ak}\frac{\Theta }{4\ell _{P}^{3}}\left( \frac{\ell
_{P}}{{\cal L}}\right) ^{2(1+\Upsilon )}{\cal F}_{11}
\end{eqnarray}%
where 
\begin{equation}
{\cal F}_{11}=\left( 1+\vartheta _{113}\left( \frac{\ell _{P}}{{\cal L}}%
\right) ^{(1+\Upsilon )}....\right).
\end{equation}
Our convention  in the sequel is to write ${\cal F}_{nl}$ so that its first term is 
$\vartheta_{nl2}$, which is a pure number not depending  either on $\ell_P$ or on ${\cal L}$.
In this way%
\begin{eqnarray}
{\rm H}_{11}^{F} &=&\frac{\Theta }{4\ell _{P}^{3}}\left( \frac{\ell _{P}}{%
{\cal L}}\right) ^{2(1+\Upsilon )}{\cal F}_{11}\int d^{3}{\vec{x}}\;i\pi ({%
\vec{x}})\;\tau _{a}\partial _{a}{\xi }(\vec{x})
\end{eqnarray}%
In order to recover the standard kinetic term we have to choose 
\begin{equation}
\Theta =4\ell _{P}^{3}\left( \frac{{\cal L}}{\ell _{P}}\right)
^{2(1+\Upsilon )},
\end{equation}
which reduces to our  choice in Ref. \cite{URRU1}, for $\Upsilon=0$.  We obtain 
\begin{equation}
{\rm H}_{11}^{F}={\cal F}_{11}\int d^{3}\vec{x}\;N(\vec{x})\;i \pi (\vec{x}%
)\tau _{a}\partial _{a}{\xi }(\vec{x})
\end{equation}%
as the final result for the kinetic term piece of the Hamiltonian

\subsubsection{ Case $n=1,l=0$}

The basic quantities are 
\begin{equation}
{\rm H}_{10}^{F}=i\int d^{3}{\vec{x}}\;\pi ({\vec{x}})\;\tau _{k}\tau
_{m}\;T_{10}^{m\;k}{\xi }(\vec{x})
\end{equation}
\begin{equation}
T_{10}^{m\;k}\;=\frac{\Theta }{4\ell _{P}^{7}}\,< s_{K}^{a}\;\epsilon
^{ijk}\epsilon ^{IJK}\langle W,\xi |\hat{A}_{a}^{m}(v)\hat{w}_{iI\Delta }(v)%
\hat{w}_{jJ\Delta }(v)|W,\xi \rangle \,>
\end{equation}
Under the scaling properties we obtain 
\begin{equation}
T_{10}^{m\;k} =\delta ^{mk}\frac{1}{{\cal L}}\left( \frac{\ell _{P}}{{\cal L}%
}\right) ^{\Upsilon }\, {\cal F}_{10}
\end{equation}
with 
\begin{equation}
{\cal F}_{10}=\left( \vartheta_{102}\;+\vartheta _{103}\;\left( \frac{\ell
_{P}}{{\cal L}}\right) ^{(1+\Upsilon )}\;+...\right)
\end{equation}%
\begin{eqnarray}
{\rm H}_{10}^{F} &=&-\frac{3}{4}{\cal F}_{10}\int d^{3}{\vec{x}}\;i\;\pi ({%
\vec{x}})\;{\xi }(\vec{x})\,\, \frac{1}{{\cal L}}\left( \frac{\ell _{P}}{%
{\cal L}}\right) ^{\Upsilon }
\end{eqnarray}

\subsection{The case $n=2$}

Here we have three possibilities: $(n=2,l=0),\;(n=2,l=1)\;$and\ $(n=2,l=2)$

\subsubsection{Case$\;(n=2,l=2)$}

\begin{equation}
{\rm H}_{22}^{F}=i\int d^{3}{\vec{x}}\;\pi ({\vec{x}})\;\tau
_{k}\;\;T_{22}^{ab\;k}\;\partial _{a}\partial _{b}{\xi }(\vec{x})
\end{equation}
\begin{equation}
T_{22}^{ab\;k}\;=\frac{\Theta }{4\ell _{P}^{7}}<s_{K}^{a}s_{K}^{b}\;\epsilon
^{ijk}\epsilon ^{IJK}\langle W,\xi |\hat{w}_{iI\Delta }(v)\hat{w}_{jJ\Delta
}(v)|W,\xi \rangle >
\end{equation}
From the above equation we see that $T_{22}^{ab\;k}$ must be symmetric in
the indices $a,b.$ Nevertheless, the only three index tensor at our disposal
in flat space is $\epsilon ^{abk}$ so we must conclude that this
contribution is zero, that is to say$\;{\rm H}_{22}^{F}=0.$

\subsubsection{Case$\;(n=2,l=1)$}

\begin{equation}
{\rm H}_{21}^{F}=2i\int d^{3}{\vec{x}}\;\pi ({\vec{x}})\;\tau _{k}\tau
_{m}T_{21}^{am\;k}\;\partial _{a}{\xi }(\vec{x})
\end{equation}
\begin{equation}
T_{21}^{am\;k}\;=\frac{\Theta }{4\ell _{P}^{7}}<s_{K}^{a}s_{K}^{b}\;\epsilon
^{ijk}\epsilon ^{IJK}\langle W,\xi |\hat{A}_{b}^{m}(v)\hat{w}_{iI\Delta }(v)%
\hat{w}_{jJ\Delta }(v)|W,\xi \rangle >.
\end{equation}
Here we have no symmetry requirement, so the antisymmetric tensor is allowed
and we have%
\begin{eqnarray}
T_{21}^{am\;k} &=&\epsilon ^{amk}\left( \frac{\ell _{P}}{{\cal L}}\right)
^{1+\Upsilon }{\cal F}_{21}, \qquad {\cal F}_{21}=\left(\vartheta_{212} +
\vartheta_{213} \left( \frac{\ell _{P}}{{\cal L}}\right) ^{1+\Upsilon
}+...\right)
\end{eqnarray}%
The final contribution is 
\begin{eqnarray}
{\rm H}_{21}^{F} &=&-2 {\cal F}_{21}\int d^{3}{\vec{x}}\;i\pi ({\vec{x}}%
)\tau _{a}\partial _{a}{\xi }(\vec{x})\,\, \left( \frac{\ell _{P}}{{\cal L}}%
\right) ^{1+\Upsilon }
\end{eqnarray}%
This is a correction of order $\left( \frac{\ell _{P}}{{\cal L}}\right)
^{1+\Upsilon }$ to the standard kinetic term.

\subsubsection{Case $n=2,$ $l=0$}

\begin{equation}
{\rm H}_{20}^{F}=i\int d^{3}{\vec{x}}\;\pi ({\vec{x}})\;\tau
_{k}\;\;T_{20}^{nm\;k}\;\tau _{n}\tau _{m}{\xi }(\vec{x})
\end{equation}
\begin{equation}
T_{20}^{nm\;k}\;=\frac{\Theta }{4\ell _{P}^{7}}<s_{K}^{a}s_{K}^{b}\;\epsilon
^{ijk}\epsilon ^{IJK}\langle W,\xi |\hat{A}_{a}^{n}(v)\hat{A}_{b}^{m}(v)%
\hat{w}_{iI\Delta }(v)\hat{w}_{jJ\Delta }(v)|W,\xi \rangle \;(v)>
\end{equation}
Here we do have symmetry requirements. Since the operators $\hat{A}%
_{a}^{n}(v)\; $and$\;\hat{A}_{b}^{m}(v)$ commute, the above tensor must be
symmetrical in the indices $n,m.$ Again, in flat space the only three index
tensor at our disposal is the $\epsilon ^{nmk},$ which leads to ${\rm H}%
_{20}^{F}=0.$

\subsection{The case $n=3$}

Here we have four possibilities: $(n=3,\;l=0),\,(n=3,\;l=1),\, (n=3,\;l=2)$
and $(n=3,\;l=3).$

\subsubsection{Case$\;(n=3,\;l=3)$}

\begin{equation}
{\rm H}_{33}^{F}=i\int d^{3}{\vec{x}}\;\pi ({\vec{x}})\;\tau
_{k}\;\;T_{33}^{abc\;k}\;\partial _{a}\partial _{b}\partial _{c}{\xi }(\vec{x%
})
\end{equation}
\begin{equation}
T_{33}^{abc\;k}\;=\frac{\Theta }{4\ell _{P}^{7}}<s_{K}^{a}s_{K}^{b}s_{K}^{c}%
\epsilon ^{ijk}\epsilon ^{IJK}\langle W,\xi |\hat{w}_{iI\Delta }(v)\hat{w}%
_{jJ\Delta }(v)|W,\xi \rangle >
\end{equation}%
Here we have to impose the symmetry in the indices $a,b,c$. In flat space
the required tensor is%
\begin{equation}
t_{33}^{abck}=\frac{1}{3}\left( \delta ^{ab}\delta ^{ck}+\delta ^{bc}\delta
^{ak}+\delta ^{ca}\delta ^{bk}\right)
\end{equation}%
which turns out to be symmetric in all four indices Then 
\begin{eqnarray}
T_{33}^{abc\;k} &=&t_{33}^{abck}\ell _{P}^{2}\;{\cal F}_{33}, \qquad {\cal F}%
_{33}= \left( \vartheta_{332} + \vartheta_{333} \left( \frac{\ell _{P}}{%
{\cal L}}\right) ^{1+\Upsilon }+...\right)
\end{eqnarray}%
Combining the above results we obtain 
\begin{eqnarray}
{\rm H}_{33}^{F} &=&\;{\cal F}_{33}\int d^{3}{\vec{x}}\;i\pi ({\vec{x}}%
)\;\tau _{a}\partial _{a}\partial ^{2}{\xi }(\vec{x})\,\, \ell _{P}^{2},
\end{eqnarray}
the leading term of which is independent of the scale ${\cal L}$.

\subsubsection{Case $n=3,\;l=2$}

\begin{equation}
{\rm H}_{32}^{F}=3\int d^{3}{\vec{x}}\;i\pi ({\vec{x}})\;\tau
_{k}\;\;T_{32}^{abm\;k}\tau _{m}\partial _{a}\partial _{b}{\xi }(\vec{x})
\end{equation}
\begin{equation}
T_{32}^{abm\;k}\;=\frac{\Theta }{4\ell _{P}^{7}}<s_{K}^{a}s_{K}^{b}s_{K}^{c}%
\;\epsilon ^{ijk}\epsilon ^{IJK}\langle W,\xi |\hat{A}_{c}^{m}(v)\hat{w}%
_{iI\Delta }(v)\hat{w}_{jJ\Delta }(v)|W,\xi \rangle >
\end{equation}

The above tensor must be symmetrical in the indices $a,b$ only. In flat
space we can construct%
\begin{equation}
t_{32}^{abmk}=\alpha _{32}\delta ^{ab}\delta ^{mk}+\beta _{32}\left( \delta
^{am}\delta ^{bk}+\delta ^{bm}\delta ^{ak}\right)
\end{equation}%
Then we have 
\begin{equation}
T_{32}^{abm\;k} =\;t_{32}^{abmk}\ell _{P}^{2}\frac{1}{{\cal L}}\left( \frac{%
\ell _{P}}{{\cal L}}\right) ^{\Upsilon }{\cal F}_{32}, \qquad {\cal F}_{32}=
\left( \vartheta _{322}+\vartheta _{323}\left( \frac{\ell _{P}}{{\cal L}}%
\right) ^{(1+\Upsilon )}+....\right)
\end{equation}
which leads to 
\begin{equation}
{\rm H}_{32}^{F}=-3\left( \frac{3}{4}\alpha _{32}+\frac{1}{2}\beta
_{32}\right) {\cal F}_{32}\int d^{3}{\vec{x}}\;i\pi ({\vec{x}})\;\ell
_{P}^{2}\partial ^{2}{\xi }(\vec{x})\;\, \,\frac{1}{{\cal L}}\left( \frac{%
\ell _{P}}{{\cal L}}\right) ^{\Upsilon }
\end{equation}

\subsubsection{Case $n=3,\;l=1$}

\begin{equation}
{\rm H}_{31}^{F}=3\int d^{3}{\vec{x}}\;i\pi ({\vec{x}})\;\tau
_{k}\;\;T_{31}^{amn\;k}\;\tau _{m}\tau _{n}\partial _{a}{\xi }(\vec{x})
\end{equation}
\begin{equation}
T_{31}^{amn\;k}\;=\frac{\Theta }{4\ell _{P}^{7}}<s_{K}^{a}s_{K}^{b}s_{K}^{c}%
\;\epsilon ^{ijk}\epsilon ^{IJK}\langle W,\xi |\hat{A}_{b}^{m}(v)\hat{A}%
_{c}^{n}(v)\hat{w}_{iI\Delta }(v)\hat{w}_{jJ\Delta }(v)|W,\xi \rangle >
\end{equation}
The above tensor must be symmetrical in $m,n$. We have%
\begin{equation}
t_{31}^{amnk}=\alpha _{31}\delta ^{mn}\delta ^{ak}+\beta _{31}\left( \delta
^{ma}\delta ^{nk}+\delta ^{na}\delta ^{mk}\right)
\end{equation}%
Then 
\begin{equation}
T_{31}^{amn\;k}=t_{31}^{amnk}\left( \frac{\ell _{P}}{{\cal L}}\right)
^{2\left( 1+\Upsilon \right) }\;{\cal F}_{31},\qquad {\cal F}_{31}= \left(
\vartheta _{312}+\vartheta _{313}\left( \frac{\ell _{P}}{{\cal L}}\right)
^{(1+\Upsilon )}+....\right)
\end{equation}
which produces

\begin{equation}
{\rm H}_{31}^{F}=-\frac{1}{4}{\cal F}_{31}\left( 9\alpha _{31}+6\beta
_{31}\right) \int d^{3}{\vec{x}}\;i\pi ({\vec{x}})\tau _{a}\partial _{a}{\xi 
}(\vec{x})\left( \frac{\ell _{P}}{{\cal L}}\right) ^{2\left( 1+\Upsilon
\right) }
\end{equation}

\subsubsection{Case $n=3,\;l=0$}

\begin{equation}
{\rm H}_{30}^{F}=\int d^{3}{\vec{x}}\;i\pi ({\vec{x}})T_{30}^{mnp\;k}\;\tau
_{k}\tau _{m}\tau _{n}\tau _{p}{\xi }(\vec{x})
\end{equation}
\begin{equation}
T_{30}^{mnp\;k}\;=\frac{\Theta }{4\ell _{P}^{7}}<s_{K}^{a}s_{K}^{b}s_{K}^{c}%
\epsilon ^{ijk}\epsilon ^{IJK}\langle W,\xi |\hat{A}_{a}^{m}(v)\hat{A}%
_{b}^{n}(v)\hat{A}_{c}^{p}(v)\hat{w}_{iI\Delta }(v)\hat{w}_{jJ\Delta
}(v)|W,\xi \rangle >
\end{equation}
The above tensor is symmetrical with respect to the indices $m,n,p$. This
means that we need 
\begin{equation}
t_{30}^{mnpk}=\frac{1}{3}\left( \delta ^{mn}\delta ^{pk}+\delta ^{np}\delta
^{mk}+\delta ^{pm}\delta ^{nk}\right)
\end{equation}%
so that 
\begin{equation}
T_{30}^{mnp\;k}=t_{30}^{mnpk}\frac{\ell _{P}^{2}}{{\cal L}^{3}}\left( \frac{%
\ell _{P}}{{\cal L}}\right) ^{3\Upsilon }\;{\cal F}_{30}, \;\qquad {\cal F}%
_{30}=\left( \vartheta _{302}+\vartheta _{303}\left( \frac{\ell _{P}}{{\cal L%
}}\right) ^{(1+\Upsilon )}+.....\right),
\end{equation}
which produces 
\begin{equation}
{\rm H}_{30}^{F} =\frac{15}{48}{\cal F}_{30}\int d^{3}{\vec{x}}\;i\pi ({\vec{%
x}})\;{\xi }(\vec{x})\,\, \frac{\ell _{P}^{2}}{{\cal L}^{3}}\left( \frac{%
\ell _{P}}{{\cal L}}\right) ^{3\Upsilon }
\end{equation}

\subsection{The full contribution ${\rm H}^F$}

It is given by 
\be
{\rm H}^F=\sum_{n,l}\, {\rm H}^F_{nl},
\ee
where the different pieces were calculated in the previous subsection. Next we make a clear separation among terms containing either $\ell_P$ or  ${\cal L}\,$  and those purely numerical factors not including these quantities, through the following series
of redefinitions. First we change ${\cal F}_{nl}$ into $ {\cal G}_{nl}$, where 
 \begin{equation}
{\cal G}_{nl}=\left( \varkappa _{nl2}+\varkappa _{nl3}\left( \frac{\ell _{P}%
}{{\cal L}}\right) ^{(1+\Upsilon )}+.....\right), 
\end{equation}%
by absorbing all numerical factors  in the latter coefficients,  in such a way that the Hamiltonian is written as 
\begin{eqnarray*}
{\rm H}^{F} &=&\int d^{3}\vec{x}\;N(\vec{x})\;i\pi (\vec{x})\left( {\cal G}%
_{11}+{\cal G}_{21}\left( \frac{\ell _{P}}{{\cal L}}\right) ^{1+\Upsilon }+%
{\cal G}_{31}\left( \frac{\ell _{P}}{{\cal L}}\right) ^{2\left( 1+\Upsilon
\right) }+{\cal G}_{33}\ell _{P}^{2}\nabla^{2}+..\right)  \\
&&\times \tau _{a}\partial _{a}{\xi }(\vec{x}) \\
&&+\int d^{3}{\vec{x}}\;\frac{i}{4}\;\pi ({\vec{x}})\;\frac{1}{{\cal L}}\left( \frac{%
\ell _{P}}{{\cal L}}\right) ^{\Upsilon }\left( {\cal G}_{10}+{\cal G}%
_{30}\left( \frac{\ell _{P}}{{\cal L}}\right) ^{2(1+\Upsilon )}+{\cal G}%
_{32}\ell _{P}^{2}\nabla ^{2}+..\right) {\xi }(\vec{x})\,\,
\end{eqnarray*}%
In particular we have $%
{\cal G}_{11}\equiv {\cal F}_{11}$ and what we have done amounts effectively  to a numerical redefinition  of  each
$\vartheta_{nlN}$ into $\varkappa_{nlN}$. Finally 
we factor out the contributions arising
from different powers of  $\left( \frac{\ell _{P}}{{\cal L}}\right)
$ and $\ell_P$. To the order considered we have 
\begin{eqnarray}
{\cal G}_{11}+{\cal G}_{21}\left( \frac{\ell _{P}}{{\cal L}}\right)
^{1+\Upsilon }+{\cal G}_{31}\left( \frac{\ell _{P}}{{\cal L}}\right)
^{2\left( 1+\Upsilon \right) }+... &=& 1 +\kappa _{1}\left( 
\frac{\ell _{P}}{{\cal L}}\right) ^{1+\Upsilon }+\kappa _{2}\left( \frac{\ell _{P}}{{\cal L}}\right) ^{2\left(
1+\Upsilon \right) }+...\nonumber \\
{\cal G}_{33}&=&\frac{\kappa_3}{2}
\end{eqnarray}
where $\kappa _{P}$ are numbers not containing either $\ell_P$ or ${\cal L}$. The factor one in the RHS of the first equation arises from the condition that in the limit $\ell_P\rightarrow0 $ we recover the standard fermionic Hamiltonian. Analogously we redefine
\begin{eqnarray}
{\cal G}_{10}+{\cal G}_{30}\left( \frac{\ell _{P}}{{\cal L}}\right)
^{2(1+\Upsilon )}&=&\kappa_4 +\kappa_{5}\left( \frac{\ell _{P}}{{\cal L}}%
\right) ^{1+\Upsilon }+\kappa_{6}\left( \frac{\ell _{P}}{{\cal L}}\right)
^{2\left( 1+\Upsilon \right) }+.......\nonumber \\
 {\cal G}_{32}&=&\frac{\kappa_7}{2}
\end{eqnarray}
where the remaining $\kappa$'s fulfill the same condition as the previous ones.
The above leads to%
\begin{eqnarray}
&&{\rm H}^{F} =\int d^{3}\vec{x}\;N(\vec{x})\;i\pi (\vec{x})\left( 1 +\kappa_{1}\left( \frac{\ell _{P}}{{\cal L}}\right) ^{1+\Upsilon
}+\kappa _{2}\left( \frac{\ell _{P}}{{\cal L}}\right) ^{2\left(
1+\Upsilon \right) }+\frac{\kappa_3}{2}\,\ell _{P}^{2}\nabla ^{2}+...\right)\tau _{a}\partial _{a}{\xi }(\vec{x}) \nonumber \\
&&+\int d^{3}{\vec{x}}\;\frac{i}{4}\;\pi ({\vec{x}})\;\frac{1}{{\cal L}}\left( \frac{%
\ell _{P}}{{\cal L}}\right) ^{\Upsilon } \left( \kappa_{4}+\kappa_{5}\left( \frac{\ell _{P}}{{\cal L}}%
\right) ^{1+\Upsilon }+\kappa_{6}\left( \frac{\ell _{P}}{{\cal L}}\right)
^{2\left( 1+\Upsilon \right) }+ \frac{\kappa_7}{2}\,\ell _{P}^{2}\nabla
^{2}+..\right) {\xi }(\vec{x})\,\,\nonumber \\
\end{eqnarray}
which is our final expression for this piece of the effective Hamiltonian.

\subsection{The mass term}

Since the spinors $\xi,\pi$ are half-densities regularization of the mass
term requires a treatment along the lines of \cite{Thiemann} that
effectively dedensitize them. For our purposes, however, it is enough to
determine the leading contribution and this can be done consistently with
our strategy above. Namely, using the triangulation again, 
\begin{eqnarray}
H_{m}&=& \frac{m}{2\hbar}\int d^3x N \xi^T( i \sigma^2)\xi + c.c.
\nonumber \\
&=& \sum_{\Delta} N(v(\Delta)) \frac{8}{3!E(v)}\epsilon^{IJK}\epsilon^{abc}
s^a_I(\Delta)s^b_J(\Delta)s^c_K(\Delta) \xi^T(s(\Delta)) \ ( i \sigma^2) \
h_{s(\Delta)} \xi(v(\Delta))+ c.c.  \nonumber \\
\end{eqnarray}
and in the quantum theory, by adapting the triangulation to the flat WBSC state
with fermions, we get 
\begin{eqnarray}
{\rm H}_m=\langle W,\xi |\hat H_{m}|W,\xi \rangle &=& \sum_{\Delta}
N(v(\Delta)) \frac{8}{3!E(v)}\epsilon^{IJK}\epsilon^{abc}
s^a_I(\Delta)s^b_J(\Delta)s^c_K(\Delta)  \nonumber \\
&& \times \langle W,\xi |\theta^T(s(\Delta))\ ( i \sigma^2)\ h_{s(\Delta)}\
\theta(v(\Delta))|W,\xi \rangle +c.c.  \nonumber \\
&=& \sum_{v\in V(\gamma)}N(v(\Delta)) \frac{8}{3!E(v)}\epsilon^{IJK}%
\epsilon^{abc} s^a_I(\Delta)s^b_J(\Delta)s^c_K(\Delta)  \nonumber \\
&&\times \left(\frac{}{} \langle W,\xi |\theta^T(v)( i \sigma^2)\theta(v)|W,\xi \rangle
\right.  \nonumber \\
&& \mbox{} + \langle W,\xi |s^a\partial_a\theta^T(v)\ ( i \sigma^2)\theta(v)|W,\xi
\rangle \nonumber \\
&&+ \langle W,\xi |\theta^T(v) \ ( i \sigma^2) \ s^aA_a(v) \ \theta(v)|W,\xi
\rangle  \nonumber \\
& & \mbox{} \left. + \langle W,\xi |s^a\partial_a\theta^T(v)\ ( i \sigma^2)\
s^aA_a(v)\ \theta(v)|W,\xi \rangle + \cdots \frac{}{}\right) +c.c.\nonumber \\
\end{eqnarray}
Hence, to leading order, the modifications coming from Planck scale to
standard flat space mass term for fermions are just 
\begin{eqnarray}
{\rm H}_m &=& \frac{m}{2\hbar}
\int d^3x \left( \xi^T( i \sigma^2) \xi + \kappa_9 \ \ell_P\, 
\xi^T \ ( i \sigma^2)\ \tau^a \, \partial_a \xi \right.  \nonumber \\
&& \left. + \kappa_{8} \, \left(\frac{\ell_P}{{\cal L}}\right)^{\Upsilon+1} \xi^T( i \sigma^2)\xi +
\kappa_{11}\,\ell_P \, \left(\frac{\ell_P}{{\cal L}} \right)^{\Upsilon+1} \xi^T \ ( i
\sigma^2) \ \tau^a \,\partial_a \xi + c.c. \right),  \label{EFFHM}
\end{eqnarray}
where we have set $N=1$.

\subsection{The extra contributions}

Here we study the second term of Eq. (\ref{eq:h12})
\begin{equation}
\label{eq:Dpixi}
H^{(2)}_{\rm spin-\frac{1}{2}}:= \int_{\Sigma} d^3x\,\frac{E^a_i}{2\sqrt{q}} [{\cal D}_a \left(\pi^T\tau_i\xi\right) + c.c.]
\end{equation}
Repeating the procedure applied to $H^{(1)}_{\rm spin-\frac{1}{2}}$ step by step with the pertinent modification in the fermionic term leads to
\begin{eqnarray}
\hat{H}^{(2)}_{\rm spin-1/2}[N] &=& - \frac{m_P}{2\ell_P^3} \sum_{v\in V(\gamma)}  N_v  \sum_{v(\Delta)=v}
\epsilon^{ijk} \epsilon^{IJK} \nonumber\\
&& \mbox{} \times {\rm tr} \left( \tau_i h_{s_I(\Delta)}\left[h^{-1}_{s_I(\Delta)}, \sqrt{\hat V_v)}\right]\right)  {\rm tr} \left( \tau_j h_{s_J(\Delta)}\left[h^{-1}_{s_J(\Delta)}, \sqrt{\hat V_v)}\right]\right)\nonumber \\
&& \mbox{} \times \left[\left[(h^{-1}_{s_K(\Delta)}\tau_k h_{s_K(\Delta)}\theta)(s_K(\Delta)(\delta))\right]_A\frac{\partial}{\partial\theta_A(s_K(\Delta)(\delta))} \right. \nonumber\\
&& \mbox{} \left. - \left[\tau_k\theta(v)\right]_A\frac{\partial}{\partial\theta_A(v)} + h.c.\right] \; .
\label{eq:h(2)}
\end{eqnarray}
The corresponding expectation value behaves as  
\begin{equation}
\langle\hat H^{(2)}_{\rm spin-\frac{1}{2}}\rangle =  \int_{\Sigma}d^3x  
[Z_1\,\partial_a B^a +  Z_2\, \partial_a {\partial_b} B^{ab} + \dots ]
\end{equation}
where the $Z_i$'s represent the estimate of the non fermionic factors in (\ref{eq:h(2)}) and the $B$'s represent the fermionic bilinears. Clearly every $B$ term, being a boundary term, will not affect the effective dynamics of the spin-$\frac{1}{2}$ particle.

We proceed to describe the third term in Eq. (\ref{eq:h12}) 
\begin{equation}
H^{(3)}_{\rm spin-\frac{1}{2}}:= 
\int_{\Sigma} d^3x\,\frac{E^a_i}{2\sqrt{q}} \frac{1}{2}[i K_a^i \pi^T\xi + c.c.] \; .
\label{eq:xK}
\end{equation} 
To regularize it we can treat the $\frac{E^a_i}{2\sqrt{q}}$ factor similarly as with $H^{(1)}_{\rm spin-\frac{1}{2}}$ term. As for the $K_a^i$ factor we recall \cite{Thiemann}
\begin{eqnarray}
K_a^j &=&-\frac{1}{\kappa}\left\{ A_a^j,\left\{V,H_E\right\} \right\} \label{eq:KVHE}\\
H^E &=& \frac{2}{\kappa} \int_{\Sigma} d^3x \,\epsilon^{abc} {\rm tr} \left(F_{ab}\left\{ A_c,V\right\}\right) \,.
\end{eqnarray}
Upon regularization of the quantum $H^E$ one gets
\begin{eqnarray}
H^E &=& \sum_{\Delta} H^E_{\Delta} \; , \\
H^E_{\Delta} &=& -\frac{2N_v}{3i\kappa\ell_P^2}\epsilon^{ijk} {\rm tr}
\left( h_{\alpha_{ij(\Delta)}} h_{s_k(\Delta)}\left[h^{-1}_{s_k(\Delta)},V_v\right]\right)\,. \label{eq:HEDelta}
\end{eqnarray}
Following the standard procedure to incorporate holonomies and fermions one arrives at the quantum version of (\ref{eq:xK}) as
\begin{eqnarray}
&& \left\{ \sum_{v\in V(\gamma)}\frac{\hbar}{\ell_P^{10}}  \sum_{v(\Delta)=v}\sum_{v(\Delta ')=v}\epsilon^{IJK}\epsilon^{LMN} \epsilon_{ijk}\;
w_{iI\Delta} w_{jJ\Delta} \right. \nonumber\\
&& \mbox{} \times  tr \left(\tau_k h_{s_K} \left[h_{s_K}^{-1},\left[V_v,  {\rm tr}
\left( h_{\alpha_{LM(\Delta ')}} h_{s_N(\Delta ') }\left[h^{-1}_{s_N(\Delta ')},V_v\right]\right) \right]\right]\right)\nonumber\\
&& \left. \mbox{}\times \xi_A(v(\Delta))\frac{\partial}{\partial \xi_A(v(\Delta))} + h.c.\right\} f_{\gamma}
\end{eqnarray}
where the prefactor $\frac{\hbar}{\ell_P^{10}}$ was obtained from $\left(\frac{1}{\kappa^2}\right)(\hbar) \left(\frac{1}{\hbar^2}\right) \left(\frac{1}{\kappa^2\hbar^2\ell_P^2}\right)$ with the $\kappa$ factors coming from the different Poisson brackets identities, the $\hbar$ factors from the quantization procedure and the $\ell_P^2$ factor arises similarly as in previous steps.
We want to estimate the leading order contribution of (\ref{eq:xK}) to the effective dynamics of the spin-$\frac{1}{2}$ particle.
For simplicity in the discussion we set $\Upsilon=0$, but an extension can be given for $\Upsilon >0$. The leading order is estimated by noticing that under $<\,\langle \dots\rangle\,>$ the contribution of an holonomy $h$ is 1 if it is not an entry of a 
commutator, otherwise it scales like $\ell_P/{\cal L}$. Such a leading order contribution behaves then like $\frac{\hbar}{\ell_P^{10}}\left(\ell_P^{3/2}\frac{\ell_P}{\cal L}\right)^2 \left(\frac{\ell_P}{\cal L}\right)^2 \ell_P^6= \frac{\hbar}{\ell_P}\left(\frac{\ell_P}{\cal L}\right)^4 = m_P\left(\frac{\ell_P}{\cal L}\right)^4$,  which is highly suppressed given the quartic order and $\ell_P<\!\!< {\cal L}$.

\section{Effective dynamics for spin-$\frac{1}{2}$ particles}

The total effective Hamiltonian ${\rm H}_T$ for our spin-$\frac{1}{2}$ particle is then
\begin{eqnarray}  \label{EFFHF}
{\rm H}_T={\rm H}^F+{\rm H}_m &=& \int d^3 x \left[ i \
\pi(\vec x) \tau^d\partial_d \ {\hat A} \right. \xi({\vec x})
+ c.c.  \nonumber \\
&+& \frac{i}{4\hbar} \frac{1}{{\cal L}} \ \pi({\vec x}) \,{\hat C}\,
\xi({\vec x})  \nonumber \\
&+& \frac{m}{2 \hbar } \xi^T({\vec x})\ (i \sigma^2)
\left( \alpha + 2\hbar\,\beta \ \tau^a
\partial_a \right)\xi({\vec x})  \nonumber \\
&+&\left. \frac{m}{2 \hbar} \pi^T({\vec x}) \left( \alpha  + 2\hbar\,\beta \  \tau^a \partial_a
\right) (i \sigma^2) \pi({\vec x}) \right],
\end{eqnarray}
where 
\begin{eqnarray}\label{EFFHF1}
&&{\hat A}=\left(1 + { \kappa}_{1} \left(\frac{\ell_P}{{\cal L}} \right)^{\Upsilon+1}+ { \kappa}%
_{2} \left(\frac{\ell_P}{{\cal L}} \right)^{2\Upsilon+2} + \frac{{ \kappa}_3}{2} \
\ell_P^2 \ \ \nabla^2 \right),\nonumber\\ 
&&{\hat C}=\hbar \ \left({ \kappa}_4 \left( \frac{\ell_P}{\cal L}\right)^\Upsilon
+ { \kappa}_{5} \left(\frac{\ell_P}{%
{\cal L}}\right)^{2\Upsilon+1} 
+ { \kappa}_{6}\left(\frac{\ell_P}{{\cal L}} \right)^{3\Upsilon+2} +\frac{%
{ \kappa}_{7}}{2} \left(\frac{\ell_P}{\cal L}\right)^{\Upsilon}\ \ell_P^2 \ \ \nabla^2\right),
\nonumber \\
&&  \alpha= \left(1 + { \kappa}_{8} 
\left(\frac{\ell_P}{{\cal L}}\right)^{\Upsilon+1}\right),
 \qquad \beta=\frac{{%
 \kappa}_9}{2\hbar} \ell_P + \frac{\kappa_{11}}{2\hbar}\ell_P\left(\frac{\ell_P}{\cal L} \right)^{\Upsilon+1},
\end{eqnarray}

It can be seen  that the terms proportional to $\kappa_4, \kappa_5$  and $\kappa_6$ correspond to a renormalization of the fermion mass. They will not be considered in the sequel, although they would give rise to standard neutrino oscillations even if the neutrino bare mass is zero.

The wave equation becomes
\begin{eqnarray}  \label{E2}
&&\left[i \hbar \frac{\partial}{\partial t}-i \ \hbar\ {\hat A }\ {\vec
\sigma} \cdot \nabla +\frac{{\hat C}}{2{\cal L}} \right]\xi(t,{\vec x}) +m
\left( \alpha -\beta \ i\hbar \ {\vec \sigma}\cdot \nabla \right)\chi(t,{%
\vec x})=0,   \\
&&\left[i \hbar \frac{\partial}{\partial t}+ i \ \hbar\ {\hat A} \ {\vec
\sigma} \cdot \nabla -\frac{{\hat C}}{2{\cal L}} \right]\chi(t,{\vec x})
+m\left( \alpha - \beta \ i\hbar \ {\vec \sigma}\cdot \nabla \right) \xi(t,{
\vec x})=0,  
\label{E21} 
\end{eqnarray}
with $\chi(t,{\vec x})=i \ \sigma_2 \xi^*(t,{\vec x})$. Following the standard steps one can
verify the consistency of the above equations. 

Eliminating $\chi$ from (\ref{E2}) we obtain 
\begin{equation}
\chi= - \frac{1}{m}\, \frac{1}{\alpha - \beta \ i \ \hbar {\vec \sigma}\cdot
\nabla } \left[i \hbar \frac{\partial}{\partial t}-i \ \hbar\ {\hat A }\ {
\vec \sigma} \cdot \nabla +\frac{{\hat C}}{2{\cal L}} \right]\xi
\end{equation}
Substituting in (\ref{E21}) we obtain the second order equation 
\begin{equation}
\left( \left[i \hbar \frac{\partial}{\partial t}+ i \ \hbar\ {\hat A} \ {%
\vec \sigma} \cdot \nabla -\frac{{\hat C}}{2{\cal L}} \right]\left[i \hbar 
\frac{\partial}{\partial t}-i \ \hbar\ {\hat A }\ {\vec \sigma} \cdot \nabla
+\frac{{\hat C}}{2{\cal L}} \right] -m^2\left( \alpha - \beta \ i \ \hbar {%
\vec \sigma}\cdot \nabla \right)^2 \right)\xi(t,{\vec x})=0.
\end{equation}
The above equation has positive and negative energy solutions 
\begin{equation}  \label{PWS}
W({\vec p},h) e^{-\frac{i}{\hbar} E t + \frac{i}{\hbar} {\vec p}\cdot {\vec x%
}}, \quad W({\vec p},h) e^{\frac{i}{\hbar} E t - \frac{i}{\hbar} {\vec p}%
\cdot {\vec x}},
\end{equation}
where it is convenient to take the spinorial part $W({\vec p},h)$ as
helicity $({\vec \sigma}\cdot {\hat p})$ eigenstates, with $h=\pm 1$, so
that 
\begin{equation}
W({\vec p},1)=\left(%
\begin{array}{c}
{\rm cos}(\frac{\theta}{2}) \\ 
e^{i \phi} \ {\rm sin}(\frac{\theta}{2})%
\end{array}%
\right), \quad W({\vec p},-1)=\left(%
\begin{array}{c}
-e^{-i \phi} \ {\rm sin}(\frac{\theta}{2}) \\ 
{\rm cos}(\frac{\theta}{2})%
\end{array}
\right).
\end{equation}

The dispersion relation has the following form 
\begin{equation}  \label{DR}
E_{\pm}=\sqrt{(A^2+ m^2 \beta^2 )|{\vec p}|{}^2+m^2 \ \alpha^2+\left(\frac{C%
}{2{\cal L}}\right)^2 \ \pm \ B\ |{\vec p}|},
\end{equation}
with 
\begin{eqnarray}
&&{ A}=\left(1 + { \kappa}_{1} \left(\frac{\ell_P}{{\cal L}} \right)^{\Upsilon+1}+ { \kappa}%
_{2} \left(\frac{\ell_P}{{\cal L}} \right)^{2\Upsilon+2} - \frac{{ \kappa}_3}{2} \
\ell_P^2 \ p^2 \right),\quad  B= A \left(\frac{C}{\cal L}+ 2\,\alpha\,\beta\, m^2 \right), \nonumber\\ 
&&{ C}= -\frac{\hbar{\kappa}_{7}}{2} \left(\frac{\ell_P}{\cal L}\right)^{\Upsilon}\ \ell_P^2 \ p^2 \, ,
\nonumber \\
&&  \alpha= \left(1 + { \kappa}_{8} 
\left(\frac{\ell_P}{{\cal L}}\right)^{\Upsilon+1}\right),
 \qquad \beta=\frac{{%
 \kappa}_9}{2\hbar} \ell_P + \frac{\kappa_{11}}{2\hbar}\ell_P\left(\frac{\ell_P}{\cal L} \right)^{\Upsilon+1},
\end{eqnarray}
The $\pm$ in Eq. (\ref{DR}) refers to the dispersion relation of the
helicities $\pm$ respectively. Let us emphasize that the solution $\xi(t,{%
\vec x})$ to either Eq.(\ref{E2}) or Eq.(\ref{E21}) is given by an appropriate
linear combination of plane waves and helicity eigenstates. This is not
unexpected since we are dealing with massive neutrinos.


In the sequel we write down the dispersion relation  in terms of an expansion up to second order in $\left(\frac{\ell_P}{{\cal L}}\right)^\Upsilon$. The coefficient of each power is subsequently expanded in powers of $m$ in the combinations either $\left(\frac{m}{p}\right)$ or $ (m\ell_P)$. In this way we obtain
\begin{eqnarray}
E_\pm(p, {\cal L}) &=&\left[ p+\frac{m^{2}}{2p}\pm \ell _{P}\left( \frac{1}{2}m^{2}\kappa
_{9}\right) +\ell _{P}^{2}\left( -\frac{1}{2}\kappa _{3}p^{3}+\frac{1}{8}%
\left( 2\kappa _{3}+\kappa _{9}^{2}\right) m^{2}p\right) \right]\nonumber  \\
&&+\left( \frac{\ell _{P}}{{\cal L}}\right) ^{\Upsilon+1 }\left[  \left( \kappa _{1}p-\frac{\Theta _{11}m^{2}}{4p}\right)  
\pm \ell_P\,\left( -\kappa _{7}\frac{p^{2}}{4}+\Theta _{12}\frac{m^{2}%
}{16}\right) \right]\nonumber  \\
&&+\left( \frac{\ell _{P}}{{\cal L}}\right) ^{2\Upsilon +2}\left( \kappa _{2}p-\frac{m^{2}}{64p}\Theta
_{22}\right)  
\label{CDR2}
\end{eqnarray}
where the previous coefficients denoted by $\kappa$ appear in the following combinations
\be
\Theta _{11}=\left( 2\kappa _{1}-4\kappa _{8}\right), \qquad 
\Theta _{12}=\left( 8\kappa _{11}+2\kappa
_{7}+8\kappa _{8}\kappa _{9}\right) , 
\ee
\begin{eqnarray}
\Theta _{22} &=&-32\kappa _{1}^{2}+32\kappa _{2} + 64\kappa _{1}\kappa _{8}
\end{eqnarray}

The velocity is defined as 
\begin{equation}
v_\pm(p, {\cal L})= \frac{\partial E_\pm(p, {\cal L})}{\partial p}.
\end{equation}
and is
\begin{eqnarray}
v_\pm(p, {\cal L})&=&\left[ \left( 1-\frac{m^{2}}{2p^{2}}%
\right) +\ell _{P}^{2}\left( -\frac{3}{2}\kappa _{3}p^{2}+\frac{1}{8}\left(
2\kappa _{3}+\kappa _{9}^{2}\right) m^{2}\right) \right] \nonumber \\
&&+\left( \frac{\ell _{P}}{{\cal L}}\right) ^{\Upsilon+1 }\left[ \left(
\kappa _{1}+\frac{\Theta _{11}m^{2}}{4p^{2}}\right)\mp\, \frac{\kappa
_{7}}{2}(\ell_P\,p) \right] \nonumber \\
&&+\left( \frac{\ell _{P}}{{\cal L}}\right) ^{2\Upsilon +2}\left( \kappa _{2}+\frac{m^{2}}{64p^{2}}\Theta
_{22}\right), 
\label{VELG}
\end{eqnarray}
within the same approximation.

We will be  mainly interested in the case of ultrarelativistic cosmological neutrinos, which mass we take to be $m=10^{-9}$GeV, in the range of momenta $10^5 \leq p \leq 10^{10}$  GeV, where
\begin{equation}
p >> m, \quad \Rightarrow\quad  (p\,\ell_P)>> (m\,\ell_P)\frac{m}{p}, \qquad (p\, \ell_P)^2 >> (m\,\ell_P)^2.  
\end{equation}
Such regime  allows us to introduce some simplifications in the coefficients of $\left( \frac{\ell_P}{\cal L}\right)^\Upsilon$. The results are
\begin{eqnarray}
\ell _{P}E_\pm(p,{\cal L})&=&\left( p\ell _{P}\right) +\left( \ell _{P}m\right) \frac{m}{%
2p}\pm \left( \frac{1}{2}\left( \ell _{P}m\right) ^{2}\kappa _{9}\right) -%
\frac{1}{2}\kappa _{3}\left( \ell _{P}p\right) ^{3} \nonumber \\
&&+\left( \frac{\ell _{P}}{{\cal L}}\right) ^{\Upsilon +1}\left[ \kappa
_{1}\left( p\ell _{P}\right)\mp \kappa _{7}\frac{%
\left( \ell _{P}p\right) ^{2}}{4}\right]+\left( \frac{\ell _{P}}{{\cal L}}\right) ^{2\Upsilon +2}\kappa _{2} (p\ell _{P}) ,
\label{gen1}
\end{eqnarray}
together with
\begin{eqnarray}
v_\pm(p,{\cal L}) &=&1-\frac{m^{2}}{2p^{2}}-\frac{3}{2}\kappa _{3}\left( \ell
_{P}p\right) ^{2}+\left( \frac{\ell _{P}}{{\cal L}}\right) ^{\Upsilon +1}\left[ \kappa
_{1} \mp \frac{\kappa _{7}}{2}\left( \ell _{P}p\right) \right] +\left( \frac{\ell _{P}}{{\cal L}}\right) ^{2\Upsilon +2}\, \kappa_{2}.
\label{gen2}
\end{eqnarray}

\section{Parameter estimates}

\subsection{The scale ${\cal L}$}

In order to estimate some  numerical values for the modifications to the velocity
of propagation we must further fix the value of the scale $\cal L$. Two distinguished cases arise

\subsubsection{The mobile scale  ${\cal L}=\lambda=\frac{1}{p}$}

Recall that { ${\cal L}$ is a
macroscopically length scale, being defined by the given geometry which
indicates where the non perturbative states of the spin-network can be
approximated by the classical flat metric}. The neutrino is characterized by
energies which probe to distances of order $\lambda$. To be described by a
classical continuous flat geometry equation, as the found Dirac equation, it
is necessary that one  follows in the correct range with respect to ${\cal L}$%
, i.e. ${\cal L}< \lambda$. We take the marginal choice of the equality in
order to be able to make some further estimates. Previous expressions reduce to 
\begin{eqnarray}
\ell _{P}E_\pm(p,{\cal L})|_{{\cal L}=\frac{1}{p}}&=&\left( p\ell _{P}\right) +\left( \ell _{P}m\right) \frac{m}{%
2p}\pm  \frac{1}{2}\left( \ell _{P}m\right) ^{2}\kappa _{9} -%
\frac{1}{2}\kappa _{3}\left( \ell _{P}p\right) ^{3} \nonumber\\
&+&\left( p\ell _{P}\right) ^{\Upsilon +1}\left[ \kappa _{1}\left( p\ell
_{P}\right) \mp \kappa _{7}\frac{\left( \ell
_{P}p\right) ^{2}}{4}\right]+\left( p\ell _{P}\right) ^{2\Upsilon +2} \kappa _{2}\left( p\ell _{P}\right) \nonumber \\
\label{ENDR}
\end{eqnarray}%
and 
\begin{eqnarray}
v_\pm(p,{\cal L})|_{{\cal L}=\frac{1}{p}}=1-\frac{m^{2}}{2p^{2}}-\frac{3}{2}\kappa _{3}\left( \ell _{P}p\right)
^{2}+\left( {\ell _{P}\,p}\right) ^{\Upsilon +1}\left[ \kappa
_{1}\mp \frac{\kappa _{7}}{2}\left( \ell _{P}p\right) \right]+\left({\ell _{P}\,p}\right) ^{2\Upsilon +2}\kappa
_{2}
\label{VELEQ}
\end{eqnarray}
In the case $\Upsilon=0$ we recover the expressions (9) and (10) of Ref.\cite{URRU1}.

\subsubsection{A universal scale for ${\cal L}$ }
Recently, in \cite{ALF} a universal value for $\cal L$ was considered in the context of the GZK anomaly. The study of the different reactions involved produces a preferred bound on ${\cal L}: \, 4.6\times 10^{-8}GeV^{-1}\geq{\cal L}\geq 8.3 \times 
10^{-9}GeV^{-1}$. Actually, in this case the general expressions (\ref{gen1}) and (\ref{gen2}) are valid.

\subsection{The exponent  $\Upsilon$}

In this section we follow the conclusion derived fom  the Super-Kamiokande 
atmospheric neutrino  experiment 
that neutrino oscillations  are well described by the mass differences contribution to the energy ($n=-0.9\pm 0.4$ at 90\% CL, in the standard notation \cite{FOGLI} )  \cite{BRUSTEIN} .That is to say, any additional contribution to the oscillation length must be highly suppressed.  This condition will set a lower bound to $\Upsilon$ which we  will use in further estimates.

The oscillation length $L=\frac{2\,\pi}{|\Delta E|}$ is
\be
\frac{1}{L}= \frac{1}{L_m}+ \frac{1}{L_{QG}},\qquad L=L_m\, \frac{1}{1+\frac{L_m}{L_{QG}}},
\qquad  L_m=\frac{4\pi\, E}{(\Delta m)^2},
\label{OSCLM}
\end{equation}
where we have used $p\approx E$,  $\ell_P=1/M_P$ and $L_{QG}$ is calculated according to each specific additional term in  $|\Delta E|$. Since the dominant contribution to the oscillation  arises form the mass term, i.e. $L\approx L_m$ we must  have the condition
\be
{L_{QG}} > {L_m} \approx X \,,
\ee
where $X$ is the distance travelled by the neutrinos in each experiment.
Let us observe that the energy range for current neutrino observatories lies between $10^{-2}$ GeV and $10^{+2}$ GeV, which amply fulfills the condition $(p\,{\cal L})\leq 1$. In particular we will consider the case of the SNO experiment characterized by the following parameters \cite{SNO}
\be
E = 10^{-2} GeV,  \quad X=10^8 km =  10^{27} \, 1/GeV
\ee
 The estimates proceed by considering first the separate contribution
of each arbitrary parameter  $\Delta \kappa$ and subsequently the case where we have more than one non-zero contribution to the oscillation.

\subsubsection{The case ${\cal L}=\lambda=\frac{1}{p}$}

(i) $\Delta \kappa_3 \neq 0,\, \Delta \kappa_1= \Delta \kappa_7 = 0 $

The quantum gravity contribution to the oscillation length is
\begin{eqnarray}
L_{QG}=\frac{4\pi}{\Delta\kappa_3}\left(\frac{M}{E}\right)^2\,\frac{1}{E},
\end{eqnarray}
from where we get that
$$
\Delta \kappa_3 < 6.28\times 10^{17},
$$
which is a very weak bound on the variable. The reason for this is 
that this term corresponds to a second order correction in $l_P$.

(ii) $\Delta \kappa_1 \neq 0,\, \Delta \kappa_3= \Delta \kappa_7 = 0 $.

Here we have
\be
L_{QG}=\frac{2\pi}{\Delta\kappa_1}\left(\frac{M}{E}\right)^{\Upsilon+1}\,\frac{1}{E},
\ee 
Using $\Delta \kappa_1 \sim 1$, we obtain
\be
\Upsilon > 0.152.
\ee

(iii) $\Delta \kappa_7 \neq 0,\, \Delta \kappa_1= \Delta \kappa_3 = 0 $.

Here we have
\be
L_{QG}=\frac{8\pi}{\Delta\kappa_7}\left(\frac{M}{E}\right)^{\Upsilon+2}\,\frac{1}{E}.
\ee 
Using $\Delta \kappa_7 \sim 1$, we obtain
\be
\Upsilon > -0.87.
\ee

(iv) When two of the $\Delta \kappa_i$ do not simultaneously vanish, we get the following  situations:

(iv-1) $\Delta \,\kappa_1 \neq 0$

In this case the term proportional to $\Delta\kappa_7$ is suppressed with respect to the one proportional to
$\Delta\kappa_1$. The comparison among the terms proportional to $\Delta\kappa_1$ and $\Delta\kappa_3$ leads to the threshold $\Upsilon=1$. When $\Upsilon  > 1$ the term proportional to $\Delta\kappa_3$ dominates, while the term proportional to $\Delta\kappa_1$ dominates in the other situation.

(iv-2) $\Delta \,\kappa_1 = 0$

In this case the competition is among the terms proportional to $\Delta\kappa_3$ and $\Delta\kappa_7$. The threshold here is $\Upsilon =0$, so that we consider only the case $\Upsilon > 0$, where the term proportional to  $\Delta\kappa_3$ dominates.

\subsubsection{A universal scale for ${\cal L}$}

In this estimate we will assume that ${\cal L}=10^{-8} Gev^{-1}$ \cite{ALF}. 

(i) $\Delta \kappa_3 \neq 0,\, \Delta \kappa_1= \Delta \kappa_7 = 0 $

Since this contribution does not depend upon ${\cal L}$, we obtain the same result as the corresponding case in the mobile
scale.

(ii) $\Delta \kappa_1 \neq 0,\, \Delta \kappa_3= \Delta \kappa_7 = 0 $.

Here we have
\be
L_{QG}=\frac{2\pi}{\Delta\kappa_1}\left(\frac{{\cal L}}{\ell_P}\right)^{\Upsilon+1}\,\frac{1}{E}.
\ee 
Using $\Delta \kappa_1 \sim 1$, we obtain
\be
\Upsilon > 1.2.
\ee

(iii) $\Delta \kappa_7 \neq 0,\, \Delta \kappa_1= \Delta \kappa_3 = 0 $.

Here we have
\be
L_{QG}=\frac{8\pi}{\Delta\kappa_7}\left(\frac{{\cal L}}{\ell_P}\right)^{\Upsilon+1}\, \frac{M_P}{E}\,\frac{1}{E}.
\ee 
Assuming $\Delta \kappa_7 \sim 1$, we get a bound on $\Upsilon$:
\be
\Upsilon >-0.764,
\ee
which is also a weak bound on the variable.

(iv) When two of the $\Delta \kappa_i$ do not simultaneously vanish, we get the following  situations:

(iv-1) $\Delta \kappa_1 \neq 0 $, which leads to the threshold $\Upsilon>\frac{17}{11}$. When $\Upsilon > 17/11$ the term $\Delta \kappa_3$ dominates and we are back to the case  (i). On the other hand, when $\Upsilon < \frac{17}{11}$ we obtain the interval 
$\frac{17}{11}>\Upsilon>1.2$. \, 
It should be stressed that the latter  bound on $\Upsilon$ is also compatible
with the bound on the maximum speed  of  Ref. \cite{coleman}, which in our case reads
\be
\Delta v_{QG}=\left|v_\pm(p, {\cal L})-\left(1-\frac{m^2}{2p^2}\right)\right |=| \kappa_1|\left( \frac{\ell _{P}}{{\cal L}}\right) ^{\Upsilon
+1}<10^{-22}. \label{dA}
\ee
(iv-2) $\Delta \kappa_1=0$

The threshold here is $\Upsilon=3/11$. When  $\Upsilon > 3/11$  the term proportional to
$\Delta \kappa_3$  dominates over the one proportional to $\Delta \kappa_7$

So, the present data on neutrino oscillations do not proscribe the  theory considered in this work  with a universal scale ${\cal L}
\sim 10^{-8}\, GeV^{-1}$

\section{Summary and Discussion}

In this work we have derived an effective Hamiltonian exhibiting Planck scale corrections with respect to  standard propagation for the theory describing spin 1/2 fermions, using an heuristical approach based upon Thiemann's regularization within the framework of  loop quantum gravity. Corrections arise as a consequence of the discrete nature of space which are manifest at Planck scale. The effective spin 1/2 particle Hamiltonian given in Eqs.(\ref{EFFHF},\ref{EFFHF1}) was obtained by taking the expectation value of the regularized version of the quantum operator corresponding to Eq.(\ref{hmax}) with respect to  a {\it would be semiclassical state} $|W, \xi \rangle$ describing a large scale flat metric together with a slowly varying  classical spinor field. Only the basic properties of such state were used: (i) peakedness both in a flat space metric together with  a flat connection for large  distances  $d >>{\cal L}>>\ell_P$, where
 ${\cal L}$ can be thought of as the scale that settles the transition between a discrete and a continuous description of space, (ii) well defined  expectation values, (iii) existence of a coarse-grained expansion involving ratios of the relevant scales of the problem: the Planck length $\ell_P$, the characteristic
length ${\cal L}$ of the state and the de Broglie wavelength $\lambda$ of the spin
1/2 particle and (iv) invariance under rotations at scales larger than ${\cal L}$, which amounts to express the box-averaged values introduced after Eq.(\ref{GWEV}) only in terms of flat space tensors.

The effective theory violates Lorentz invariance and, in analogy with the photon case
\cite{FOT}, we assume that the effective Hamiltonian so found corresponds to that in the particular frame of reference where the cosmic radiation background  looks isotropic.  

Some improvements with respect to our original presentation \cite{URRU1} are the following: (i) in section 5.6 we have elucidated the contribution of the derivative term and the extrinsic curvature dependent term in Eq.(3) by showing  that they are   highly suppressed in powers of $\ell_P$. (ii) we have extended the corrections to the scaling of the connection by including the new parameter $\Upsilon$, already considered in our discussion of photons \cite{FOT}, in the form
\be
\label{IMPROV}
\langle W, \, \xi|\, A_{ia}\, |W,\, \xi\rangle=0+ \frac{1}{{\cal L}}\left(\frac{\ell_P}{{\cal L}}\right)^\Upsilon,
\ee  
where $\Upsilon >0 $ can be any real number. 

In Section 8 we have estimated some bounds for  $\Upsilon$, based on the observation that atmospheric neutrino oscillations at average energies of the order $10^{-2}-10^2 $ GeV are dominated by the corresponding mass differences via the oscillation length $L_m$ in Eq. (\ref{OSCLM}). This means that additional contributions to the oscillation length, in particular the quantum gravity correction $L_{QG}$,  should satisfy $L_{QG}> L_m$, which is used to set a lower bound upon
$\Upsilon.$ Within the proposed two different ways of estimating the scale ${\cal L}$ of the process we obtain: (i) $\Upsilon > 0.15$ when ${\cal L}$ is considered as a mobile scale and it is  estimated by $1/E$ and (ii) $1.2 <\Upsilon $ when the scale ${\cal L}$ takes the universal value ${\cal L}=10^{-8}\,  1/GeV$, according to  Ref. \cite{ALF}.  In the above estimates we have satisfied the condition $(p\, {\cal L})\leq 1$.

Let us observe that according to Eq.(\ref{ENDR}) the mass-difference contribution to the neutrino oscillation process will be highly suppressed at high energies while other mechanisms, like those arising in the quantum gravity framework, could be the dominant ones \cite{SOUTHHAMPTON}. It is an experimental issue to settle this question. Since the bound upon $\Upsilon$ strongly depends on the dominant mechanism there is  the possibility that the exponent $\Upsilon$ be energy-dependent. This situation is not considered in the present approach. Nevertheless, 
in order to make some numerical estimates related to cosmological neutrinos we will make the extrapolation of our bounds in $\Upsilon$, which have been obtained  in the range of a few GeV, to energies of the order of $10^5$\, GeV ,  corresponding to typical neutrinos  arising from Gamma Ray Bursts. Let us consider specifically ultrarelativistic neutrinos of mass $m=10^{-9} $ GeV, energy $p \approx E= 10^5 $ GeV, traveling a cosmological distance  $L=10^{10} l.y. =
0.5\times 10^{42} $ 1/GeV, where $p{\cal L}\leq 1$. Next we give an  estimate of two types of delay times (for a more realistic calculation including the effect of the expanding universe see Refs.\cite{FOT}, \cite{ALF},
\cite{SOUTHHAMPTON}): 

(i) ${\Delta t}_\nu \approx \frac{L}{c^2}%
\left(v-v_0\right)$ , which measures the time delay of the arrival of
the neutrino coming from a distance $L$, flying with velocity $v=\frac{%
\partial E}{\partial p}$, with respect to the time of flight corresponding
to a particle of mass $m$ with  velocity  $v_0=\frac{p}{\sqrt{p^2+ m^2}}\approx 1$.

(ii) ${\Delta t}_{\pm}\approx \frac{%
L}{c^2} \left(v_- - v_+ \right)
 $, which is a measure of the birrefringence effects and it  is defined  by the difference in
the arrival time of two neutrinos with opposite polarizations 

In the range $0.6 < \Upsilon < 2.0 $ and within the two scenarios presented here for dealing with the scale ${\cal L}$, we obtain the estimates
\ba
10^{-10}\, {\rm s} &<&\Delta t_\nu \, <\, 10^{-4}\, s, \\
&& \Delta t_\pm \,< \, 10^{-17} {\rm s}. 
\ea

To conclude we point out that an effective dynamics and dispersion relations for gravity plus matter were thoroughly studied recently by Thiemann and Sahlmann \cite{THIESAHL}, where the semiclassical states were taken as coherent states. They included photons and scalar particles and their results for the dispersion relations essentially agree with what we obtain here for fermions,  even the possibility of having non integral powers in $\ell_P$ for the correcting terms, which is encoded in our parameter $\Upsilon$. Besides they provide a detailed classification of the quantum geometry aspects required in defining the semiclassical regime yielding the effective dynamics. Yet another avenue to tackle the problem of defining semiclassical states in quantum geometry is currently under investigation that establishes a relation between Fock space and the kinematical Hilbert space for diffeomorphism covariant theories of connections such as quantum geometry/loop quantum gravity
\cite{FockU1,Fockpolymer} (see also \cite{madzap}). It will be interesting to compare all of the above proposed semiclassical states in the context of the quantum gravity phenomenology.

\

\

{\bf Acknowledgments}. The authors would like to thank A. Ashtekar, M. Bojowald, R.
Gambini and T. Thiemann for suggestions on this work.
Partial support is acknowledged from: the bilateral program CONICyT-CONACyT, DGAPA IN11700 and
CONACYT 32431-E. LFU acknowledges the hospitality of CERN together with support from the program CERN-CONACyT (M\'exico). We also acknowledge the project Fondecyt 7010967.
 The work of JA is partially supported by Fondecyt 1010967. He acknowledges the hospitality 
of LPTENS (Paris) and CERN; and financial support from an Ecos(France)-Conicyt(Chile) project. HAMT thanks for the warm hospitality and
stimulating environment experienced at the Center for Gravitational Physics and
Geometry-PSU. He also acknowledges
partial support from NSF Grants PHY00-90091 and  INT97-22514, Eberly Research Funds at Penn State,
CONACyT 55751 and M\'exico-USA Foundation for Science.


\newpage


\begin{thebibliography}{99}


\bibitem{HUET}P. Huet and M. Peskin, Nucl Phys. {\bf B434}, 3 (1995); J. Ellis, J. L\'opez, N.E. Mavromatos and D.V. Nanopoulos, Phys. Rev. {D53}, 3846 (1996).


\bibitem{AC} G. Amelino-Camelia, J. Ellis, N.E. Mavromatos, D.V. Nanopoulos
and S. Sarkar, {
Nature}\, {\bf 393} \,(1998)\, 763-765.

\bibitem{GP} R. Gambini and J. Pullin,
{\em Phys. Rev.} {\bf D59} (1999) 124021, [gr-qc/9809038].

\bibitem{URRU1} J. Alfaro, H.A. Morales-T\'ecotl and L.F. Urrutia, {Phys. Rev. Letts.} {\bf 84}(2000)2318, [gr-qc/9909079].

\bibitem{FOT}
J. Alfaro, H.A. Morales-T\'ecotl and L.F. Urrutia, Phys. Rev. {\bf D65} 103509 (2002), [hep-th/0108061].

\bibitem{gleiser}
R.J. Gleiser and C.N. Kozameh, 
Phys. Rev. {\bf D64} 083007 (2001), [gr-qc/0102093].


\bibitem{ACLFLUC} G. Amelino-Camelia, Nature {\bf 398} (1999) 216 [gr-qc/9808029]; 
Nature {\bf 410} (2001) 1065 [gr-qc/0104086].

\bibitem{NGVDAMLFLUC} Y.J. Ng and H. van Dam, Found. Phys. {\bf 30} (2000) 795 [gr-qc/9906003].  

\bibitem{BRUSTEIN} R. Brustein, D. Eichler and S. Foffa,  Phys. Rev. {\bf 65}, 105006 (2002).


\bibitem{KIFUNE} T. Kifune, Astrophys. J. Lett. {\bf 518} (1999) 
L21-L24 [astro-ph/9904164].

\bibitem{ACPIRAN} G. Amelino-Camelia and T. Piran, Phys. Rev. {\bf D64} (2001) 036005 [gr-qc/0008107].

\bibitem{AMC4} G. Amelino-Camelia, Phys.Lett.{\bf B 528} 181-187 (2002), [gr-qc/0107086].


\bibitem{ALF} J. Alfaro and G. Palma, Phys.Rev.{\bf D65} (2002)103516. [hep-th/0111176].


\bibitem{CPTELLIS} J.R. Ellis, J.L. Lopez, N.E. Mavromatos and D.V. Nanopoulos,
Phys. Rev. {\bf D53} (1996) 3846-3870 [hep-ph/9505340].

\bibitem{SUDVUUR}D. Sudarsky, L.F. Urrutia and H. Vucetich, {\it New Observational bounds to quantum gravity signals}, [gr-qc/0204027]. 

\bibitem{ELLIS} J. Ellis, N.E. Mavromatos and D.V. Nanopoulos,
{ Gen. Rel. Grav.} {\bf 31} (1999) 1257 [gr-qc/9905048].

\bibitem{AHLU} D.V. Ahluwalia, 
{ Nature} {\bf 398} (1999) 199; G.Z. Adunas, E. Rodriguez-Milla and D.V. Ahluwalia,
 { Phys. Letts.} {\bf B485}(2000)215-223,
[gr-qc/0006021]; G.Z. Adunas, E. Rodriguez-Milla and D.V. Ahluwalia, { Gen. Rel. Grav.} {\bf
33}(2001)183-194,[gr-qc/0006022]; D.V. Ahluwalia, C.A. Ortiz and G.Z. Adunas, {
Robust flavor equalization of cosmic neutrino flux by quasi bimaximal mixing},
[hep-ph/0006092].

\bibitem{ACPol}
G. Amelino-Camelia,  { Are we at the dawn of quantum gravity phenomenology?},
Lectures given at 35th Winter School of Theoretical Physics: From Cosmology to
Quantum Gravity, Polanica, Poland, 2-12 Feb 1999. Published in Lect. Notes Phys. 541
(2000) 1-49, Springer Verlag  [gr-qc/9910089].

\bibitem{ACEssay}
G. Amelino-Camelia, { Int. Jour. Mod. Phys.} {\bf D10}
(2001) 1, [gr-qc/0008010].

\bibitem{ACINDIA}
G. Amelino-Camelia, Quantum-gravity phenomenology: status and prospects [gr\-qc/0204051].

\bibitem{THIESAHL}  H. Sahlmann and T. Thiemann, {\it Towards the QFT on curved spacetime limit
of QRG. I: A general scheme}, [gr-qc/0207030]; H. Sahlmann and T. Thiemann, {\it Towards the QFT on curved spacetime limit
of QRG.II: A concrete implementation}, [gr-qc/0207031].


\bibitem{METZ} J. van Paradis et. al., { Nature} {\bf 386}(1997)686; P.J.
Groot et al.,IAU Circ. No 6676, 1997; M. L. Metzger et al.,  {
Nature} {\bf 387 }(1997)878.

\bibitem{BHAT} P.N. Bhat, G.J. Fishman, C.A. Meegan, R.B. Wilson, M.N. Brock
and W.S. Paclesas, {
Nature} {\bf 359}(1992)217.

\bibitem{meszaros} P. M\'esz\'aros, { Nucl. Phys. B (Proc. Suppl.)} {\bf 80} (2000) 63-77.

\bibitem{BILLER} S.D. Biller et. al., Phys. Rev. Letts. {\bf 83}, 2108 (1999).


\bibitem{WAX} E. Waxman and J. Bahcall, { Phys. Rev. Letts.} {\bf 78}(1997) 2292-2295; E. Waxman, { Nucl. Phys. (Proc. Supl)} {\bf 91} (2000)
494-500 ; { Nucl. Phys. (Proc. Supl)}
{\bf 87} (2000) 345-354.

\bibitem{VIETRI} M. Vietri, 
 { Phys. Rev. Letts.} {\bf 80}(1998) 3690-3693.

\bibitem{ROY} M. Roy, H.J. Crawford and A. Trattner,  The prediction and
detection of UHE Neutrino Bursts, [astro-ph/9903231].

\bibitem{cline} D.B. Cline and F.W. Stecker, Exploring the ultrahigh
energy neutrino universe, [astro-ph/0003459].

\bibitem{halzen} F. Halzen,
 { Phys. Rep.} {\bf 333} (2000) 349-364.


\bibitem{SOUTHHAMPTON} S. Choubey and  S.F. King, {\it Gamma ray bursts as probes of neutrino mass, quantum gravity and dark energy}, arXiv: hep-ph/0207260 .

\bibitem{REVIEW} N.E. Mavromatos, { The quest for quantum gravity:
testing times for theories?}, [astro-ph/0004225]; J. Ellis,{ Perspectives in
High-Energy Physics}, JHEP Proceedings, SILAFAE III, Cartagena de Indias, Colombia,
april 2-8,2000, [hep-ph/0007161]; J. Ellis, {\it Testing fundamental physics with
high-energy cosmic rays}, astro-ph/0010474.


\bibitem{ELLISETAL} J. Ellis, N.E. Mavromatos and D.V. Nanopoulos,  {\em Gen. Rel. Grav.}
{\bf 32} (2000) 127-144, [gr-qc/9904068]; J. Ellis, N.E. Mavromatos and D.V.
Nanopoulos, Tegernsee 1999, Beyond the desert, p.299-334,[gr-qc/9909085] and
references therein; J. Ellis, K. Farakos, N.E. Mavromatos, V.A. Mitsou and D.V.
Nanopoulos, {\em
Astrophysical Jour.} {\bf 535} (2000) 139-151, [astro-ph/9907340].

\bibitem{ELLISFERM} J. Ellis, N.E. Mavromatos, D.V. Nanopoulos and G.
Volkov, {Gen.
Rel. Grav.} {\bf 32} (2000) 1777-1798, [gr-qc/9911055].


\bibitem{LAMBIASE} G. Lambiase,{Gen.
Rel. Grav.} {\bf 33}, 2151 (2001),[gr-qc/0107066]; Eur. Phys. J. {\bf C19}, 553 (2001).

\bibitem{opensystems}
F. Benatti and  R. Floreanini, Massless neutrino oscillations,  [hep-ph/0105303];
 { Phys. Rev.} {\bf D62} (2000)
125009,    [hep-ph/0009283].

\bibitem{oneloop-effqg}
D.A.R. Dalvit, F.D. Mazzitelli, C. Molina-Paris,  { Phys. Rev.} {\bf D63}(2001) 084023, [hep-th/0010229].

\bibitem{jacobson}
T. Jacobson and D. Mattingly, Phys. Rev. {\bf D63} (2001) 041502(R);
T. Jacobson, Lorentz violation and Hawking radiation, [gr-qc/0110079].

\bibitem{MAJOR} T.J. Konopka and  S.A. Major, 
New J. Phys. {\bf 4} (2002)57. [hep-ph/0201184].

\bibitem{colladay} D. Colladay and V.A. Kostelecky, Phys. Rev. {\bf D55}, 6760 (1997);  Phys. Rev. {\bf D58} (1998) 116002 [hep-ph/9809521];
V.A. Kostelecky and C.D. Lane, J. Math. Phys. {\bf 40},6245 (1999); V.A. Kostelecky and  R. Lehnert, Phys. Rev. {\bf D63}, 065008 (2001); D. Colladay and P. McDonald, [hep-ph/0202066].   For a recent review see for example V.A. Kostelecky, Topics in Lorentz and CPT violation, [hep-ph/0104227] and references therein.

\bibitem{BLUHM} R. Bluhm, [hep-ph/0111323] and references therein. 


\bibitem{tritium}
J.M. Carmona and  J.L. Cort\'es,  Phys. Lett. B {\bf 494}(2000)75-80 [hep-ph/0007057]; Infrared and
ultraviolet cutoffs of quantum field theory [hep-th/0012028].

\bibitem{liberati}
S. Liberati, T. Jacobson and  D. Mattingly, High energy constraints on Lorentz symmetry violations
[hep-ph/0110094]; T. Jacobson, Stefano Liberati and D. Mattingly,
TeV astrophysics constraints on Planck scale Lorentz violation, [hep-ph/0112207].

\bibitem{PADDY} T. Padmanabhan, Phys. Rev. {D57}, 6206 (1998); K. Srinivasan, L. Sriramkumar and T. Padmanabhan, Phys. Rev. {D58}, 044009 (1998). S. Shankaranarayanan and  T. Padmanabhan, Int. J. Mod. Phys. {\bf D10}, 351 (2001).  

\bibitem{SMOLIN} G. Amelino-Camelia, Int. J. Mod. Phys. {\bf D11}, 35 (2002); J. Magueijo and L. Smolin, Phys. Rev. Letts. {\bf 88}, 190403 (2002),
[hep-th/0112090];  Generalized Lorentz invariant with an invariant energy scale, 
[gr-qc/0207085]; S. Judes and M. Visser, Conservation laws in doubly special relativity,[gr-qc/0205067];
M. Arzano and G. Amelino-Camelia, Dirac spinors for doubly special relativity, 
[gr-qc/0207003]; D.V. Ahluwalia and  M. Kirchbach, Fermions, bosons and locality in special relativity with two invariant scales, [gr-qc/0207004]. 


\bibitem{volumeop}

C. Rovelli and L. Smolin, { Nucl.
Phys.} {\bf B442} (1995) 593-622; Erratum-ibid {\bf B456} (1995) 753. A. Ashtekar and
J. Lewandowsky,  { Class. Quant. Grav.}
{\bf 14} (1997) A55-A82; { Adv.
Theor. Math. Phys.} {\bf 1} (1998) 388-429; A. Ashtekar, A. Corichi and J. Zapata,
{ Class.
Quant. Grav.} {\bf 15} (1998) 2955-2972.

\bibitem{CanonicalBlackHole}
C. Rovelli, Helv. Phys. Acta {\bf 69}
(1996) 582-611 [gr-qc/9608032]; C. Rovelli,  Phys. Rev. Lett.{\bf 77} (1996) 3288-3291,1996 [gr-qc/9603063]; A. Ashtekar,
J. Baez, A. Corichi and K. Krasnov, Phys.
Rev. Lett. {\bf 80} (1998) 904-907 [gr-qc/9710007].

\bibitem{canonicalnosingular}

M. Bojowald, Phys. Rev. Lett. {\bf
86} (2001) 5227-5230 [gr-qc/0102069].

\bibitem{RROV} For a recent review see for example C. Rovelli, Loop quantum gravity,
Livings Reviews Vol 1, 1998-1, http://www.livingreviews.org /Ar\-ti\-cles. See also
R. Gambini and J. Pullin, Loops, Knots, Gauge Theories and Quantum Gravity, Cambridge
University Press, Cambridge UK 1996.

\bibitem{Ash} A. Ashtekar, 
{ Phys. Rev. Letts.} {\bf 57} (1987) {2244}; C. Beetle and A. Corichi, { Bibliography
of publications related to classical and quantum gravity in terms of connections and
loop variables } [gr-qc/9703044].

\bibitem{qdiffgauge}
A. Ashtekar, J. Lewandowski, D. Marolf, J. Mourao and T. Thiemann, J.
Math. Phys. {\bf 36} (1995) 6456-6493 [gr-qc/9504018].


\bibitem{Barbero} F. Barbero,  {Phys. Rev. } {\bf D51} (1995) 5507-5510; { Phys. Rev. } {\bf D51} (1995) {5498-5506}.


\bibitem{WICK} T. Thiemann, { Class. Quant. Grav.} {\bf 13}
(1996) 1383-1404 [gr-qc/9511057]; A. Ashtekar, 
{ Phys. Rev.} {\bf D53} (1996) 2865-2869 [gr-qc/9511083]; A. Ashtekar, J.
Lewandowski, D. Marolf, J. Mourao, T. Thiemann,  { J. Funct. Anal.} {\bf 135} (1996) 519-551 [gr-qc/9412014].

\bibitem{REALITY} H.A. Morales-T\'ecotl, L.F. Urrutia, J.D. Vergara, { Class. Quant. Grav.} {\bf 13} (1996)
2933-2940 [gr-qc/9607044];  M. Montesinos, H.A. Morales-Tecotl, L.F. Urrutia and J.D.
Vergara, {J. Math. Phys.} {\bf 40} (1999) 1504-1517 [gr-qc/9903043]; { Gen. Rel. Grav.} {\bf 31} (1999) 719-723.

\bibitem{Thiemann} T. Thiemann,{\em Class. Quant. Grav.} {\bf 15} (1998) 1281-1314;
 {\em Class. Quant. Grav.} {\bf 15} (1998) 839-873.

\bibitem{weave} A. Ashtekar, C. Rovelli and L. Smolin,  {\em Phys. Rev. Letts.} {\bf 69}(1992) 237-240,
[hep-th/9203079]; J. Zegwaard, {\em Phys. Lett.}
{\bf B300} (1993) 217-222 [hep-th/9210033]; R. Borissov, {\em Phys. Rev.} {\bf D49} (1994) 923-929. J. Iwasaki and C.
Rovelli, {\it Int. J. Mod. Phys.} {\bf D\ 1}
(1993) 533-557; {\em Class. and Quant. Grav.} {\bf 11}
(1994) 1653-1676; J. Iwasaki, {\it Basis states for gravitons in nonperturbative loop
representation space} [gr-qc/9807013].

\bibitem{intersecting} B. Brugmann, R. Gambini and J. Pullin, {\em Phys. Rev. Letts.} {\bf 68}
(1992) 431-434.



\bibitem{twocircles}
N. Grot and C. Rovelli, {\em Gen. Rel. Grav.}
{\bf 29} (1997) 1039-1048.

\bibitem{gaussweave}
A. Corichi and J.M. Reyes, 
Int. J. Mod. Phys. D {\bf 10} (2001) 325-338 [gr-qc/0006067].

\bibitem{aei} T. Thiemann,
{\em Class. Quant. Grav.} {\bf 18} (2001) 2025-2064, [hep-th/0005233]; T. Thiemann and
O. Winkler, {\em
Class. Quant. Grav.} {\bf 18} (2001) 2561-2636, [hep-th/0005237]; T. Thiemann and O.
Winkler, Gauge field theory coherent states (GCS): III. Ehrenfest theorems
[hep-th/0005234]; Gauge field theory coherent states (GCS): IV. Infinite tensor
product and thermodynamical limit [hep-th/0005235]; H. Sahlmann, T. Thiemann, O.
Winkler, { Coherent States for Canonical Quantum General Relativity and the Infinite
Tensor Product Extension} [gr-qc/0102038].

\bibitem{statg} A. Ashtekar and L. Bombelli, Statistical geometry of
quantum spin networks: flat space, in preparation;
L. Bombelli, Statistical geometry of random weave states, [gr-qc/0101080].

\bibitem{FOGLI}  G.L. Fogli, E. Lisi, A. Marrone and G. Scioscia, Phys. Rev. {\bf D60}, 053006(1999).

\bibitem{SNO} SNO Collaboration, Q.R. Ahmad et. al., {\it Measurement of charged current interactions produced by ${}^{8} B$ solar neutrinos at Sudbury Neutrino Ob\-ser\-va\-to\-ry}; http://owl.phy.queensu.ca/sno/.

\bibitem{coleman} S. Coleman and S.L. Glashow, Phys. Rev.
D59,116008(1999).



\bibitem{FockU1}
M. Varadarajan,  Phys. Rev. {\bf
D61} (2000) 104001 [gr-qc/0001050]; Photons from quantized electric flux
representations [gr-qc/0104051].

\bibitem{Fockpolymer}
A. Ashtekar and J. Lewandowski,
Class.Quant.Grav. {\bf 18}L117-L128(2001), [gr-qc/0107043]; A. Ashtekar, S. Fairhurst, J.L. Willis, Quantum gravity, shadow states, and quantum mechanics, [gr-qc/0207106].

\bibitem{madzap}
M. Varadarajan and  J.A. Zapata, Class. Quant. Grav. {\bf 17} (2000) 4085-4110
[gr-qc/0001040].



\end{thebibliography}
\end{document}